\documentclass[10pt,twocolumn,amsmath,amssymb,aps,floatfix,preprintnumbers,nofootinbib,nobibnotes,bibnotes]{revtex4-2}
\usepackage{graphicx}% Include figure files
\usepackage{dcolumn}% Align table columns on decimal point
\usepackage{bm}% bold math
\usepackage{hyperref}% add hypertext capabilities
\usepackage[mathlines]{lineno}% Enable numbering of text and display math
\usepackage{epsfig}  
\usepackage{color}
\usepackage{slashed}
\usepackage{float}
\usepackage[table]{xcolor}

\large

\begin{document}

\title{New physics in $B \to K^* \tau^+ \tau^-$: A model independent analysis}

\author{Neetu Raj Singh Chundawat}
\email{chundawat.1@iitj.ac.in}
\affiliation{Indian Institute of Technology Jodhpur, Jodhpur 342037, India}

\begin{abstract}
In this work we consider the implications of current $b \to s \ell^+ \ell^-$ ($\ell=e,\,\mu$) measurements on several $B \to K^* \tau^+ \tau^-$ observables under the assumption that the possible new physics can have both universal as well as nonuniversal couplings to leptons.  For new physics solutions which provide a good fit to all $b \to s \ell^+ \ell^-$ data, we intend to identify observables with large deviations from the Standard Model (SM) predictions as well as to discriminate between various  new physics solutions.   For this we consider  the $B \to K^* \tau^+ \tau^-$ branching fraction, the $K^*$ longitudinal fraction $f_L$, the tau forward-backward asymmetry $A_{FB}$ and the optimized observables in the $P_i^{(')}$ basis. Further, we construct  the $\tau - \mu$ lepton-flavor differences ($Q^{\tau\mu}$) between these tau observables and their muonic counterparts in $B \to K^* \mu^+ \mu^-$ decay. Moreover, we also consider 
 lepton-flavor ratios ($R^{\tau\mu}$)   of all of these observables. We find that the current data allows for deviations ranging from 25\% up to an order of magnitude from the SM value  in a number of observables. For e.g.,  the magnitudes of $Q^{\tau \mu}_{P_3}$ and $Q^{\tau \mu}_{P'_8}$ observables can be enhanced up to an order of magnitude, a twofold enhancement in $Q^{\tau\mu}_{A_{FB}}$ is possible along with $\sim$50\% enhancement in $R^{\tau\mu}_{K^*}$ and $\sim$25\% in $R^{\tau\mu}_{A_{FB}}$. Moreover, the branching ratio of $B \to K^* \tau^+ \tau^-$ can be suppressed up to 25\%. A precise measurement of these observables can also discriminate between a number of new physics solutions.  
\end{abstract}
 
\maketitle 

\newpage

 %%%%%%%%%%%%%%%%%%%%%%%%%%%%%%%%%%%%
\section{Introduction} 
%%%%%%%%%%%%%%%%%%%%%%%%%%%%%%%%%%%%

The measurements of several observables in  decays induced by the quark level transition $b \to s \ell^+ \ell^-$($\ell=e,\, \mu$) show propitious signatures of new physics beyond the Standard Model (SM) interactions \cite{London:2021lfn}. These observables are mainly related to $B \to K^* \, \mu^+\,\mu^-$ and $B_s \to \phi\, \mu^+\,\mu^-$  decay modes. The measured value of the  optimized angular observable  $P'_5$ in $B \to K^* \, \mu^+\,\mu^-$  in 4.0 $\mathrm{GeV}^2 \le q^2 \le$ 6.0 $\mathrm{GeV}^2$ bin deviates from the SM prediction at the level of 3$\sigma$ \cite{Kstarlhcb1,Kstarlhcb2,LHCb:2020lmf,sm-angular}. Further, the  experimental value of the  branching ratio of $B_s \to \phi\, \mu^+\,\mu^-$ decay nonconcur with the SM prediction at the level of  3.5$\sigma$ level \cite{bsphilhc2,bsphilhc3}. Moreover, there is dissimilitude in the measured and SM prediction of the  branching ratio $B_s \to \mu^+ \mu^-$ decay \cite{LHCb:2021awg,Combination,ATLAS:2018cur,CMS:2019bbr,LHCb:2017rmj,Altmannshofer:2021qrr}. However, the recent measurement by the CMS Collaboration using the full Run 2 dataset \cite{CMS:2022dbz} shifts the world average of the branching ratio $B_s \to \mu^+ \mu^-$ \cite{HFLAV:2022pwe} to a value which is in excellent agreement with its SM prediction \cite{Bobeth:2013uxa,UTfit:2022hsi}. 
 These disparities can be accommodated by assuming new physics  with imperative couplings to muons.

The mismatch  between the electron and muon sector, owing to new physics beyond the SM, can be delineated by the ratio observables $R$. These observables enmesh agglomeration  between different lepton-flavors. The lepton-flavor universality violating  (LFUV) ratios  $R_{K} \equiv \Gamma(B^+ \to K^+ \mu^+ \mu^-)/\Gamma(B^+ \to K^+ e^+ e^-)$ and  $R_{K^*} \equiv \Gamma(B^0 \to K^{*0} \mu^+ \mu^-)/\Gamma(B^0 \to K^{*0} e^+ e^-)$ has been measured by the LHCb Collaboration. The measurement of $R_K$ showed a exiguous of 3.1$\sigma$ as compared to the SM value  $\approx 1$ \cite{Hiller:2003js,Bordone:2016gaq} in the 1.1 $\mathrm{GeV}^2 \le q^2 \le$ 6.0 $\mathrm{GeV}^2$ bin \cite{LHCb:2021trn}.  The measurements of  $R_{K^*}$, in the 0.045 $ \rm{GeV}^2 \le q^2 \le$ 1.1 $\rm{GeV}^2$ and 1.1 $\rm{GeV}^2 \le q^2 \le$ 6.0 $\rm{GeV}^2$ bins also had dissent with the SM at level of $\sim$ 2.5$\sigma$ \cite{rkstar}. These anomalous measurements required new physics with nonuniversal couplings to muons and electrons.

The deviation  of 3.1$\sigma$ was obtained assuming $R_K^{\rm SM}\approx 1$ which is valid only under the approximation of neglecting the QED corrections. These corrections can be significant as the lepton masses break lepton-flavor  universality (LFU) and their scales are  different from that of $b$ mass scale. However in \cite{Isidori:2020acz,Isidori:2022bzw,Nabeebaccus:2022pje} it was shown that these corrections are small. In particular the  hard-collinear logs are absent in the  structure-dependent QED corrections. Hence $R_K$ can be considered as a clean theoretical observable. The same conclusion applies to the $R_{K^*}$ observable.

A few more lepton-flavor universality  violation (LFUV)  ratios have been measured by the LHCb Collaboration. These are in $B^0 \to K_S \mu^+ \mu^-$ and $B^+ \to K^{*+} \mu^+ \mu^-$ decay modes \cite{LHCb:2021lvy}. The measured values of the ratios $R_{K_S} \equiv \Gamma(B^0 \to K_S \mu^+ \mu^-)/\Gamma(B^0 \to K_S e^+ e^-)$ and $R_{K^{*+}} \equiv \Gamma(B^+ \to K^{*+} \mu^+ \mu^-)/\Gamma(B^+ \to K^{*+} e^+ e^-)$ have relatively large errors as compared to $R_K$ and $R_{K^*}$ and are consistent with their SM predictions below 2$\sigma$ level.  Apart from these ratios, the LFUV new physics can also be captured by constructing additional LFUV observables by taking difference of optimized observables in the muon and electron sector, $Q_{P_i^{(')}}^{\mu e}=P_i^{(')\mu}-P_i^{(')e}$ \cite{Capdevila:2016ivx}. The $Q_{P'_4}^{\mu e}$ and $Q_{P'_5}^{\mu e}$ observables have been measured by the Belle Collaboration \cite{Belle:2016fev}. However, due to large errors, the measured values are consistent with the SM prediction of $\approx 0$.

 The above  $b \to s \ell \ell$ anomalies can be be analyzed in a model independent framework using the language of effective field theory. However, there can be a number of approaches under which the global analysis of  $b \to s \ell \ell$  data can be performed. The most common framework is where new physics is assumed to be present only in $b  \to s\, \mu^+\,\mu^-$ decay \cite{Descotes-Genon:2013wba,Altmannshofer:2013foa,Hurth:2013ssa,Alok:2017jgr,Alok:2019ufo,Altmannshofer:2021qrr,Carvunis:2021jga,Geng:2021nhg,Alguero:2021anc,Hurth:2021nsi,Angelescu:2021lln,Alok:2022pjb,SinghChundawat:2022zdf,Ciuchini:2021smi}. In another approach, new physics is allowed to be present in electron sector along with  muons with nonequal Wilson coefficients (WCs) \cite{Hurth:2021nsi,Hurth:2022lnw,Kumar:2019qbv}.
Thus in both of these approaches, the new physics couplings are nonuniversal in nature. 

In \cite{Alguero:2018nvb} a new  approach was explored where apart from having nonuniversal WCs affecting only $b \to s \mu^+ \mu^-$ decay, one can also have universal WCs  equally affecting all $b \to s \ell \ell$ processes, $\ell=e,\,\mu,\,\tau$.  Although LFUV new physics contributions are mandatory to explain
$R_{K^{(*)}}$ anomalies, a universal  new physics contribution which is the same for all leptons
is not ruled out even though  measurements in $b \to s e e$ sector are consistent with SM predictions. In fact, such a contribution gives rise to scenarios with a statistical significance at least as relevant
as that of only LFUV framework and can also motivate the construction of new models beyond SM including not only LFUV but also LFU new physics contributions \cite{Alguero:2018nvb,Alguero:2019ptt,Alguero:2021anc}.

 Very recently, on December 20, 2022, the LHCb Collaboration updated the measurements of $R_K$ and $R_{K^*}$ \cite{LHCb:2022qnv,LHCb:2022zom} by including the experimental systematic effects which were absent in the previous analysis. The updated measurements of $R_K$ and $R_{K^*}$ are now consistent with the SM predictions. A global analysis of  $b \to s \ell \ell$ data using the updated measurements  was performed in \cite{Ciuchini:2022wbq}.
 
 If both universal and nonuniversal new physics WCs are present then the universal couplings will generate new physics effects in $b \to s \tau^+ \tau^-$ decay with WCs in $\mu$ and $\tau$ sectors being the same. Therefore it is interesting to see what impact the universal coupling determined by the updated $b  \to s\, \ell^+\,\ell^-$ ($\ell=e,\,\mu$) data will have on observables in decays induced by the quark level transition  $b \to s \tau^+ \tau^-$. 
 In this work we study this implication for several $CP$-conserving observables in $B \to K^* \tau^+ \tau^-$ {\footnote{A few examples of correlating current B anomalies with new physics in  $b \to s \tau^+ \tau^-$ in specific model dependent scenarios can be seen in \cite{Capdevila:2017iqn,Alguero:2022wkd,deGiorgi:2022vup,Belle-II:2018jsg}.}}. Apart from analysing the branching ratio, the $K^*$ polarization fraction, the $\tau$ forward-backward asymmetry  and optimized angular observables in  $B \to K^* \tau^+ \tau^-$ decay, we also consider $\tau - \mu$ lepton-flavor differences ($Q^{\tau\mu}$) and ratios ($R^{\tau\mu}$) of these observables.

 The plan of the work is as follows. In the next section, we discuss the fit results based on the assumption  that both universal and nonuniversal coupling to leptons are present. In Sec. \ref{obs}, we provide definitions and theoretical expressions of observables related to $B \to K^* \tau^+ \tau^-$ decay used in our analysis. In Sec. \ref{results}, we provide results obtained in our work. We finally conclude in Sec.~\ref{concl}.

 %%%%%%%%%%%%%%%%%
\section{New Physics in $b \to s \mu^+ \mu^-$: LFU conserving and violating approach}
\label{lfu}
 %%%%%%%%%%%%%%%%%
In this section we discuss constraints on new physics couplings under the assumption that LFU new physics is allowed in addition to LFUV new physics contributions affecting only $b \to s \mu^+ \mu^-$ transition. Within  the SM, the effective Hamiltonian for $ b\to s \ell^+  \ell^- $ transition can be written as
\begin{widetext}
\begin{eqnarray}
\mathcal{H}^{\rm SM}_{\rm eff} &=& - \frac{\alpha_{em} G_F}{\sqrt{2} \pi} V_{ts}^* V_{tb}  \Big[ 2 \frac{C_7^{\rm eff}}{q^2}
 [\overline{s} \sigma^{\mu \nu} q_\nu (m_s P_L  + m_b P_R)b ] \bar{\ell} \gamma_\mu \ell + C_9^{\rm eff} (\overline{s} \gamma^{\mu} P_L b)(\overline{\ell} \gamma_{\mu} \ell) \nonumber\\ && + C_{10} (\overline{s} \gamma^{\mu} P_L b)(\overline{\ell} \gamma_{\mu} \gamma_5 \ell) \Big] + H.c. ,
\end{eqnarray}
\end{widetext} 
where $\alpha_{em}$ is the fine-structure constant, $G_F$ is the Fermi constant, $V_{ts}$ and $V_{tb}$ are the Cabibbo-Kobayashi-Maskawa (CKM) matrix elements and $P_{L,R} = (1 \mp \gamma_{5})/2$ are the chiral projection operators. The $q$ in the
$C_7$ term is the momentum of the off shell photon in the effective $b \to s \gamma^*$ transition. 

Assuming new physics in the form of vector  and axial-vector  operators, the new physics effective Hamiltonian for $b \to s \ell^+ \ell^-$ process can be written as 
\begin{eqnarray}
\mathcal{H}^{\rm NP}_{\rm eff} &=& -\frac{\alpha_{\rm em} G_F}{\sqrt{2} \pi} V_{ts}^* V_{tb} \left[ C_{9\ell} (\overline{s} \gamma^{\mu} P_L b)(\overline{\ell} \gamma_{\mu} \ell) \right. \nonumber \\
& & \left. + C_{10\ell} (\overline{s} \gamma^{\mu} P_L b)(\overline{\ell} \gamma_{\mu} \gamma_5 \ell)  + C^{\prime}_{9\ell} (\overline{s} \gamma^{\mu} P_R b)(\overline{\ell} \gamma_{\mu} \ell)\right. \nonumber \\
& & \left. + C^{\prime}_{10\ell} (\overline{s} \gamma^{\mu} P_R b)(\overline{\ell} \gamma_{\mu} \gamma_5 \ell)\right] + H.c.  \,\,.
\label{HNP}
\end{eqnarray} 
Here $C_{(9,10)\ell}$ and $C^{\prime}_{(9,10)\ell}$ are the new physics WCs. In the presence of LFU new physics, the WCs can be written as 
\begin{eqnarray}
C_{(9,10)e}&=&C_{(9,10)\tau}=C_{(9,10)}^U\,,\nonumber\\
C_{(9,10)e}^{\prime}&=&C_{(9,10)\tau}^{\prime}=C_{(9,10)}^{\prime U}\,,\nonumber\\
C_{(9,10)\mu}&=& C_{(9,10)}^U +C_{(9,10)\mu}^V\,,\nonumber\\
C_{(9,10)\mu}^{\prime}&=& C_{(9,10)}^{\prime U} +C_{(9,10)\mu}^{\prime V}\,. 
\end{eqnarray}
Hence the $C^U$ and $C^{\prime U}$ WCs contribute equally to all decays induced by the $b \to s \ell^+ \ell^-$ transitions whereas $C^V$ and $C^{\prime V}$ contribute only to $b \to s \mu^+ \mu^-$.

A global fit to all $b \to s \ell^+ \ell^-$ data, i.e., LFUV observables along with $b \to s \mu^+ \mu^-$ and $b \to s e^+ e^-$ observables, preferred new physics in $C_{9\mu}$ whereas a fit to all LFUV observables preferred $C_{9\mu}=-C_{10\mu}$ new physics scenario. Therefore a 2D scenario with $C_9^U$ along with $C_9^V=-C_{10}^V$ is expected to provide a good fit to data, with a significance at least at the level of fits with only LFUV contributions, if not better. This motivated  a new  approach in which apart from having nonuniversal WCs affecting only $b \to s \mu^+ \mu^-$ decay, an universal component equally affecting all $b \to s \ell \ell$ processes, $\ell=e,\,\mu,\,\tau$ was explored \cite{Alguero:2018nvb}.  Indeed such a scenario provided an extremely good fit to all $b \to s \ell^+ \ell^-$ data \cite{Alguero:2021anc}.  A complete set of all such scenarios are illustrated in Tab.~\ref{scenarios}. These scenarios can be classified into two categories:
\begin{itemize}
\item Class-A  (scenarios characterized by $C_9^U$ contribution): Here we have four favored scenarios. These solutions were first identified in \cite{Alguero:2018nvb}. In the notation of \cite{Alguero:2021anc} except S-V, all are 2D solutions.

\item Class-B  (scenarios characterized either by $C_{10}^U$ or $C_{10}^{\prime U}$  contribution): Here again we have four favored solutions. These additional scenarios which can arise naturally in several new physics models were introduced in \cite{Alguero:2019ptt}. For e.g., S-IX can be generated in 2HDM models \cite{Crivellin:2019dun} whereas other solutions can be induced in $Z'$ models with vector-like quarks \cite{Bobeth:2016llm,Crivellin:2018yvo}. 
\end{itemize}

In Tab.~\ref{scenarios}, we list the 1$\sigma$ range of WCs along with pull for each favored scenarios as obtained in \cite{Alguero:2021anc}. These values were obtained by performing a global fit to  all available data except measurements related to $\Lambda_b \to \Lambda \mu \mu$. In Tab.~\ref{scenarios}, we also provide our fit results using the methodology adopted in  refs \cite{Alok:2022pjb} and using the updated measurements of $R_K$ and $R_{K^*}$ by the LHCb Collaboration in December, 2022 \cite{LHCb:2022qnv,LHCb:2022zom}. We also use the modified world average of the branching ratio of $B_s \to \mu^+ \mu^-$ in the light of updated measurement from the CMS Collaboration \cite{HFLAV:2022pwe,CMS:2022dbz}.  In addition, we included observables from $b \to s e^+ e^-$ sector. For comparison, we also provide our fit results using data before December 2022 updates. In the following we list all observable used in our fit.

{\rowcolors{2}{pink!65!white!30}{yellow!10!white!30}
\begin{table*}[hbt]
\addtolength{\tabcolsep}{-1pt}
\begin{center}
\begin{tabular}{|c|c|c|c||c|c|c|c|}
  \hline\hline
Solutions	 & WCs & 1$\sigma$ range \cite{Alguero:2021anc} & pull \cite{Alguero:2021anc} & 1$\sigma$ range (old)& pull & 1$\sigma$ range (new) & pull \\
  \hline
S-V  & $C^{V}_{9\mu}$ & (-1.02, -0.11) &  &(-0.98,0.003) &  & (-1.31, -0.53 ) &  \\ 
     & $C^{V}_{10\mu}$ &(0.08, 0.84) & 6.6 & (0.15,0.97) & 7.7 & (-0.66 ,0.07) & 4.5 \\
     & $C^{U}_9 = C^{U}_{10}$ &  (-0.73, 0.07)  & &   (-0.76,0.08) & & (-0.13 ,0.58) &  \\
\hline
S-VI  & $C^{V}_{9\mu} = - C^{V}_{10\mu}$ & (-0.59,-0.44)  &   &(-0.60,-0.45)& & (-0.33,-0.20) &\\ 
     & $C^{U}_9 = C^{U}_{10}$ &(-0.56,-0.26) & 6.9 &(-0.44,-0.18)&7.7 & (-0.43, -0.17) & 4.1\\
 \hline
S-VII  & $C^{V}_{9\mu}$ & (-1.07,-0.63)  & & (-1.15,-0.77)& & (-0.43,  -0.08) &\\ 
     & $C^{U}_9$ &(-0.52,0.01) &6.7  & (-0.35,0.15)&7.4  & (-1.07, -0.58) & 5.5 \\
\hline
S-VIII  & $C^{V}_{9\mu} = - C^{V}_{10\mu}$  &(-0.41,-0.27) & & (-0.47,-0.32)&& (-0.18,  -0.05) & \\ 
     & $C^{U}_9$ &(-0.99,-0.63)    &7.2 &(-0.87,-0.45) & 7.9  & (-1.15,-0.77) & 5.6\\
\hline
\hline
S-IX  & $C^{V}_{9\mu} = - C^{V}_{10\mu}$ &(-0.63,-0.43) &&  (-0.61, -0.43) &&(-0.27,-0.12)&\\ 
     & $C^{U}_{10}$  &(-0.44,-0.05)    & 6.3 &(-0.32,0.07) & 7.4  & (-0.09,0.27) & 3.6\\
\hline
S-X  & $C^{V}_{9\mu}$ &(-1.13,-0.84) &&(-1.10,-0.82)& & (-0.72,-0.41) &\\ 
     & $C^{U}_{10}$   &(0.13,0.42)   & 6.9&(0.19,0.50) &7.8  & (0.05,0.34) & 4.6\\
\hline
S-XI  & $C^{V}_{9\mu}$ & (-1.20,-0.91)&&(-1.23,-0.95)& & (-0.82, -0.51) &\\ 
     & $C^{\prime U}_{10}$  &(-0.35,-0.10)    & 6.9 & (-0.37,-0.16)& 7.8  & (-0.26,-0.04) & 4.6\\
\hline
S-XIII  & $C^{V}_{9\mu}$ &(-1.27,-0.96) &&(-1.27,-0.98)& & (-0.96,-0.60) &\\ 
      & $C^{\prime V}_{9\mu}$ &(0.13,0.60) &&(0.20,0.59)& & (0.22,0.63) &\\ 
   & $C^{ U}_{10}$ &(0.10,0.47) &&(0.14,0.52)& & (0.01,0.38) &\\ 
   & $C^{\prime U}_{10}$ &(-0.15,0.21) &6.7&(-0.17,0.14)& 8.1 & (-0.08,0.24) & 5.1\\ 
 \hline
\end{tabular}
\caption{Allowed new physics solutions \cite{Alguero:2021anc}. In our fit, the pull is defined as $\sqrt{\chi^2_{\rm SM}-\chi^2_{\rm bf}}$ where $\chi^2_{\rm bf}$ is the  $\chi^2$  at the best-fit value in the presence of new physics and $\chi^2_{\rm SM}$ is the value of $\chi^2$ in the SM. The value of $\chi^2_{\rm SM}$ is $\approx 217$ before December 2022. This includes the measurements of $R_K$ and $R_{K^*}$ by the LHCb Collaboration \cite{LHCb:2021trn,rkstar}, the measurements of  $R_{K^*}$ by the Belle Collaboration \cite{Belle:2019oag} and the world average of the branching ratio of $B_s \to \mu^+ \mu^-$ as obtained in \cite{Altmannshofer:2021qrr} (denoted by `old').  In the updated fit (denoted by `new'), using the observables  listed in Sec. \ref{lfu}, the value of  $\chi^2_{\rm SM}$ reduces to $\approx 184$.}
\label{scenarios}
\end{center}
\end{table*}

We include following LFUV observables in the fit:
\begin{enumerate} 
\item The  updated measurement of $R_K$ and $R_{K^*}$ by the LHCb Collaboration in December 2022 using the full Run 1 and 2 dataset \cite{LHCb:2022qnv,LHCb:2022zom}.

\end{enumerate}
 We do not  include measurements  of the ratios $R_{K_S ^0}$ and $R_{K^{*+}}$ by the LHCb Collaboration \cite{LHCb:2021lvy} in our updated fits as they are expected to suffer from the same experimental systematic effects that lead to the updated values of  $R_K$ and $R_{K^*}$ which are now consistent with their SM predictions.  Further, we do not include $Q_{P'_4}^{\mu e}$ and $Q_{P'_5}^{\mu e}$ in the fits.

We now list $b \to s \mu^+ \mu^-$  observables used in the fit.
Here we do not include any measurements in the intermediate $q^2 $ regions, i.e., $6\, {\rm GeV}^2 \le q^2 \le 14.0 \, {\rm GeV}^2$.
\begin{enumerate}
\item   The updated world average of the branching ratio of  $B_s \to \mu^+ \mu^-$ which is $( 3.45 \pm 0.29 ) \times 10 ^{-9} $ as obtained by the HFLAV averaging group \cite{HFLAV:2022pwe,Ciuchini:2022wbq}. This updated value is due to new analysis performed by the CMS Collaboration with the full Run
2 dataset \cite{CMS:2022dbz}.

\item Recently updated differential branching fraction measurements of $B_s \to \phi \mu^+ \mu^-$ by LHCb
in various  $q^2$ intervals \cite{bsphilhc3}.  

\item The differential branching ratios of $B^0 \to K^{*0} \mu^+ \mu^- $ \cite{LHCb:2016ykl,CDFupdate,Khachatryan:2015isa}, $B^{+} \to K^{*+}\mu^{+}\mu^{-}$, $B^{0}\to K^{0} \mu^{+}\mu^{-}$and  $B^{+}\rightarrow K^{+}\mu^{+}\mu^{-}$ \cite{Aaij:2014pli,CDFupdate} in different $q^2$ bins. 

\item The branching fraction of inclusive decay mode $B \to X_{s}\mu^{+}\mu^{-}$ \cite{Lees:2013nxa} where $X_{s}$ is a hadron containing only one kaon is included in the fit in the low and high-$q^2$ bins. 

\item The longitudinal polarization fraction $f_L$, forward-backward asymmetry $A_{FB}$  and observables  $S_3$, $S_4$, $S_5$, $S_7$, $S_8$, $S_9$ in the decay $B^0 \to K^{*0} \mu^+\mu^-$
in various intervals of $q^2$, as measured by the LHCb Collaboration in 2020 \cite{LHCb:2020lmf}, along with their experimental correlations.

\item   The angular observables $f_L$, $P_1$, $P'_4$, $P'_5$, $P'_6$ and  $P'_8$ in $B^0 \to K^{*0} \mu^+\mu^-$ decay measured by ATLAS \cite{ATLAS:2018gqc} and $P_1$, $P'_5$ measured by CMS \cite{CMS:2017rzx}. 
The measurements of $f_L$ and  $A_{FB}$ by the CDF and CMS Collaborations are also included \cite{ CDFupdate,Khachatryan:2015isa}. 

\item 
The full set of angular observables for $B^+ \to K^{*+} \mu^+\mu^-$ decay mode was determined for the first time by LHCb in 2020 \cite{LHCb:2020gog}.  Here, we consider  results for $F_L$ and $P_1 - P_8 '$ optimized angular observables, along with their experimental correlation \cite{LHCb:2020gog}.

\item We include the $CP$-averaged observables $f_L$, $S_3$, $S_4$ and $S_7$ in $B_s \to \phi \mu^+\mu^-$ decays measured by the LHCb in 2021 with the available experimental correlations \cite{LHCb:2021xxq}.

\end{enumerate}

The eleven $b \to s e^+ e^-$ observables used in our global analysis are as follows:
\begin{enumerate}
\item The  measurement of differential branching fraction of $B^0 \to K^{*0}  e^+ e^-$ in $0.001 \le q^2 \le 1.0 ~{\rm GeV}^2$  bin by the LHCb Collaboration \cite{LHCb:2013pra}.

\item The  measurement of differential branching fraction of $B^+ \to K^{+} e^+ e^-$ in $1.0 \le q^2 \le 6.0 ~{\rm GeV}^2$  bin by the LHCb Collaboration \cite{LHCb:2014vgu}.

\item The measured values of the branching ratios of $B \to X_s e^+ e^-$  by the BaBar Collaboration in both low as well as high-$q^2$ bins \cite{Lees:2013nxa}.

\item The longitudinal polarization fraction $f_L$ in the decay $B^0 \to K^{*0} e^+ e^-$ in $0.002 \le q^2 \le 1.12 ~{\rm GeV}^2$  bin as measured by the LHCb Collaboration \cite{LHCb:2015ycz}.
 
\item The angular optimized observables $P'_4$ and $P'_5$ measured by the Belle Collaboration in $0.1 \le q^2 \le 4 ~{\rm GeV}^2$, $1.0 \le q^2 \le 6.0 ~{\rm GeV}^2$ and $14.18 \le q^2 \le 19.0 ~{\rm GeV}^2$ bins \cite{Belle:2016fev}.

\end{enumerate}

A global fit to above data is performed using CERN minimization code {\tt MINUIT} \cite{James:1975dr}. The  $\chi^2$ function is defined as
\begin{equation}
  \chi^2(C_i,C_j) = \big[\mathcal{O}_{\rm th}(C_i,C_j) -\mathcal{O}_{\rm exp}\big]^T \,
  \mathcal{C}^{-1} \, \big[\mathcal{O}_{\rm th}(C_i,C_j) -\mathcal{O}_{\rm exp} \big]\,,
  \label{chisq-bsmumu}
\end{equation} 
where $\mathcal{O}_{\rm th}(C_i,C_j)$ are the theoretical predictions of N  observables (179 after December 2022 update) used in the $\chi^2$  fit,  $\mathcal{O}_{\rm exp}$ are the corresponding central values of the experimental measurements and  $\mathcal{C}$ is the total covariance matrix. This  $N \times N$ matrix is constructed by adding the individual theoretical and experimental covariance matrices.
The experimental correlations are included for the angular observables in $B^0 \to K^{*0} \mu^+ \mu^-$ \cite{LHCb:2020lmf}, $B^+ \to K^{*+} \mu^+ \mu^-$ \cite{LHCb:2020gog}
and $B_s \to \phi \mu^+ \mu^-$ \cite{LHCb:2021xxq} whereas the theoretical covariance matrix includes form factors and power corrections uncertainties. These are computed using {\tt flavio} \cite{Straub:2018kue} where the observables are preimplemented based on refs. \cite{Bharucha:2015bzk,Gubernari:2018wyi}.    

We quantify the goodness of fit by pull which is defined as $\sqrt{\chi^2_{\rm SM}-\chi^2_{\rm bf}}$. Here  $\chi^2_{\rm SM}$ is the value of  $\chi^2$ in the SM and $\chi^2_{\rm bf}$ is the  $\chi^2$  at the best-fit value in the presence of new physics. We see from Tab.~\ref{scenarios} that  the value of $\chi^2_{\rm SM}$ decreased from 217 to 184 indicating that the discrepancy of data from the SM has reduced considerably. All previously favored  scenarios  still remains the favored ones but with smaller values of pull. This is expected as the overall tension between the experimental measurements and SM has reduced.  In the Sec.~\ref{results}, we obtain predictions for $B \to K^* \tau^+ \tau^-$ observables using our updated values of WCs as given in Tab.~\ref{scenarios}.  In next section we provide a description of 
these observables.

 %%%%%%%%%%%%%%%%%
 \section{$B \to K^* \ell^+ \ell^-$ observables}
\label{obs}
 %%%%%%%%%%%%%%%%%%%%%%%%%%%%
 The angular distribution of $\bar{B^0} \to \bar{K^{*0}}(\to K^-\pi^-)\ell^+\ell^-$ decay is completely encapsulated by four in dependent kinematical variables. These are  traditionally chosen to be the three angles ($\theta_{K}$, $\theta_{\ell}$ and $\phi$) and the invariant mass squared of the dilepton system ($q^2 = (p_B-p_{K^*})^2$). In the notation of ref.~\cite{LHCb:2013zuf}, the full angular decay distribution of $\bar{B^0} \to \bar{K^{*0}}(\to K^-\pi^+)\ell^+\ell^-$ decay is given by
 \begin{widetext}
\begin{equation}
\frac{d^4\Gamma}{dq^2d\cos\theta_{\ell}d\cos\theta_{K}d\phi} = \frac{9}{32\pi}I(q^2,\theta_{\ell},\theta_{K},\phi),
\end{equation}
where
\begin{eqnarray}
I(q^2,\theta_{\ell},\theta_{K},\phi) &=& I^s_1\sin^2\theta_{K} + I^c_1\cos^2\theta_{K}+(I^s_2\sin^2\theta_{K}+I^c_2\cos^2\theta_{K})\cos 2\theta_{\ell} \nonumber\\
& & +I_3\sin^2\theta_{K}\sin^2\theta_{\ell}\cos 2\phi +I_4\sin 2\theta_{K}\sin 2\theta_{\ell}\cos\phi \nonumber\\
& & + I_5 \sin 2\theta_{K}\sin\theta_{\ell}\cos\phi \nonumber \\
& & + (I^s_6\sin^2\theta_{K}+I^c_6\cos^2\theta_{K})\cos\theta_{\ell} + I_7\sin 2\theta_{K}\sin\theta_{\ell}\sin\phi \nonumber\\
& & + I_8\sin 2\theta_{K}\sin 2 \theta_{\ell} \sin\phi +I_9\sin^2\theta_{K}\sin^2\theta_{\ell}\sin 2\phi .
\label{Ifunc}
\end{eqnarray}
\end{widetext}
The twelve $q^2$ dependent angular coefficients $I^{(a)}_i$ \cite{Kruger:1999xa,Altmannshofer:2008dz,Gratrex:2015hna}
are bilinear combinations of the $K^{*0}$ decay amplitudes which in turn are functions of WCs and the form-factors which depend on the long-distance effects. The functional dependence of the angular coefficients $I^{(a)}_i$ in terms of decay amplitudes ($A_i$) are given in Appendix~(\ref{appen}).

The full angular distribution of the $CP$-conjugated mode is given by $B^0 \to K^{*0}(\to K^+\pi^-)\ell^+\ell^-$
 \begin{equation}
\frac{d^4\bar{\Gamma}}{dq^2d\cos\theta_{\ell}d\cos\theta_{K}d\phi} = \frac{9}{32\pi}\bar{I}(q^2,\theta_{\ell},\theta_{K},\phi)\,.
\end{equation}
For $\bar{B^0} \to \bar{K^{*0}}(\to K^-\pi^+)\ell^+\ell^-$ decay,  $\theta_{K}$ is the angle between the directions of kaon in the $\bar{K^{*0}}$ rest frame and the  $\bar{K^{*0}}$ in the rest frame of $\bar{B}$.  The angle $\theta_{\ell}$ is  between the directions of the $\ell^-$ in the dilepton rest frame and the dilepton in the rest frame of $\bar{B}$ whereas the angle $\phi$ is the azimuthal angle between  the plane containing the dilepton pair and the plane encompassing the kaon and pion from the $\bar{K^{*0}}$. For $B^0 \to K^{*0}(\to K^+\pi^-)\ell^+\ell^-$ decay mode,  $\theta_{K}$ is the angle between the directions of kaon in the ${K^{*0}}$ rest frame and the  ${K^{*0}}$ in the rest frame of ${B}$ whereas the angle $\theta_{\ell}$ is  between the directions of the $\ell^+$ in the dilepton rest frame and the dilepton in the rest frame of ${B}$.
 This leads to the following transformation of angular coefficients under $CP$ \cite{Belle-II:2018jsg}
\begin{equation}
I^{(a)}_{1,2,3,4,5,6} \Longrightarrow \bar{I}^{(a)}_{1,2,3,4,5,6}, \quad I^{(a)}_{7,8,9} \Longrightarrow -\bar{I}^{(a)}_{7,8,9},
\end{equation}
where $\bar{I}^{(a)}_i$ are the complex conjugate of $I^{(a)}_i$. Therefore, combining $B^0$ and $\bar{B^0}$ decays, one can construct following angular observables which depend upon the average of the distribution of the $B^0$ and $\bar{B^0}$ \cite{Altmannshofer:2008dz}
\begin{equation}
S^{(a)}_i = \frac{I^{(a)}_i(q^2)+ \bar{I}^{(a)}_i(q^2)}{d(\Gamma +\bar{\Gamma})/dq^2}.
\end{equation}
The difference of these angular coefficients will result in corresponding $CP$-violating angular observables \cite{Bobeth:2008ij,Altmannshofer:2008dz}. 

Several well-established observables in the decay of $B \to K^* \ell^+ \ell^-$ can be expressed in terms of angular coefficients 
$I^{(a)}_i$ as well as $CP$-averaged angular observables $S^{(a)}$:
\begin{itemize}
\item The angular integrated differential decay rate can be written in terms of angular coefficients as
\begin{eqnarray}
\frac{d\Gamma}{dq^2}&=&\int d\cos \theta_{\ell}\, d\cos \theta_K \,d\phi \frac{d^4\Gamma}{dq^2\, d\cos \theta_K \,d\cos \theta_{\ell}\, d\phi}\nonumber\\ &=&\frac{3}{4}\left(2I^s_1+I^c_1\right) - \frac{1}{4}\left(2I_2^3+I_2^c\right)\,.
\end{eqnarray}

\item The normalized forward-backward asymmetry can be expressed in terms of $CP$-averaged angular observables $S^{(a)}$ as
\begin{eqnarray}
A_{FB}&=& \left[\int_0^1-\int_{-1}^1\right] d\cos \theta_{\ell} \frac{d^2(\Gamma -\bar{\Gamma})}{dq^2d\cos \theta_{\ell}}/\frac{d(\Gamma +\bar{\Gamma})}{dq^2}
\nonumber\\ &=&\frac{3}{8}\left(2S^s_6+S^c_6\right)\,.
\end{eqnarray}

\item The $K^*$ longitudinal polarization fraction can be written in terms of $S^{(a)}$ observables as
\begin{eqnarray}
f_L = -S^c_2\,.
\end{eqnarray}

\end{itemize}

The $S^{(a)}$ observables are more prone to hadronic uncertainties. One can construct optimized observables with reduced uncertainties by proper combination of $f_L$ and $S^{(a)}$ observables. These observables have been proposed by several groups, see for e.g., \cite{Kruger:2005ep,Egede:2008uy,Bobeth:2011gi,Becirevic:2011bp,Matias:2012xw,Descotes-Genon:2013vna,Descotes-Genon:2013wba}. A frequently used form is the set of observables given in \cite{Descotes-Genon:2013vna,Descotes-Genon:2013wba}. A generalized and extensive analysis of angular distribution formalism can be found in ref.\cite{Gratrex:2015hna}. In this work, for $B \to K^* \tau^+ \tau^-$ decay, we consider the following set of optimized observables $P^{(')}_i$  defined in ref. \cite{Descotes-Genon:2013vna,Descotes-Genon:2013wba} and written in the basis of \cite{Kstarlhcb2}:
\begin{eqnarray}
P_1 &=& \frac{S_3}{2 S_2^s}, \,\, P_2=\frac{S_6^s}{8 S_2^s}, \,\, P_3=\frac{S_{9}}{4 S_2^s}\,,\nonumber\\
P'_4  &=& \frac{S_4}{2\sqrt{-S_2^s S_2^c}}, \,\,P'_5  = \frac{S_5}{2\sqrt{-S_2^s S_2^c}}\,,\nonumber\\
P'_6  &=& \frac{S_7}{2\sqrt{-S_2^s S_2^c}}, \,\, P'_8 = \frac{S_8}{2\sqrt{-S_2^s S_2^c}}\,.
\end{eqnarray}

The $B \to K^* \ell^+ \ell^-$ observables suffer from hadronic uncertainties which is mainly due to form factors \cite{Khodjamirian:2010vf,Bharucha:2015bzk,Gubernari:2018wyi} and nonlocal contributions associated with charm-quark loops \cite{Beneke:2001at,Khodjamirian:2010vf,Descotes-Genon:2014uoa,Capdevila:2017ert,Bobeth:2017vxj,Blake:2017fyh,Gubernari:2020eft}. The form factors in the low-$q^2$ region are calculated using  light-cone sum rules (LCSR)  or light-meson distribution
amplitudes whereas in the high-$q^2$ region, form factors are obtained from lattice computations \cite{Horgan:2013hoa,Flynn:2015ynk}.

%%%%%%%%%%%%%%%%%
 \section{Results and Discussions}
\label{results}
In this section we provide predictions for several observables in $B \to K^* \tau^+ \tau^-$ within the SM as well as for various new physics scenarios considered in Sec.~\ref{lfu}. The aim is to look for  deviations from the SM as well as to distinguish between various allowed beyond SM scenarios. The observables are classified into three categories:
\begin{itemize}
\item $\tau $ observables 
\item $\tau - \mu$ lepton-flavor differences 
\item $\tau - \mu$ lepton-flavor ratios
\end{itemize}

\begin{table*}[bt]
\centering
\resizebox{\textwidth}{!}{ 
\begin{tabular}{|c|c||c|c|c|c||c|c|c|c|c|c|} \hline
Observable & SM  & {S-V}  & {S-VI} & {S-VII} & {S-VIII}& {S-IX}  & {S-X} & {S-XI} & {S-XIII} \\ 
\hline 
$(B_{\tau\tau}^{K^*}) \times 10^{8}$	 & $2.41 \pm 0.33$  & (2.34, 2.83)     & (2.18, 2.32)    & (1.56, 1.92)    &  (1.50, 1.77) & (2.33, 2.45)&(2.31, 2.40) & (2.35, 2.41) & (2.29, 2.48)\\ 
\hline
$f_L$	 &  $0.10 \pm 0.01$   & (0.09, 0.10)   & (0.11, 0.11)  &(0.11, 0.12)    &(0.11, 0.12) & (0.10, 0.10) & (0.09, 0.10)& (0.10, 0.10)& (0.09, 0.10) \\ 
\hline
$A_{FB}$	 &  $0.21 \pm 0.02$ & (0.18, 0.22)   & (0.22, 0.23)   &(0.22, 0.23)      &(0.23, 0.23) & (0.21, 0.22) & (0.20, 0.21)&(0.21, 0.22) & (0.20, 0.21)\\ 
\hline
$P_1$	 &  $ -0.66 \pm  0.05$  &  (-0.66, -0.66) &(-0.66, -0.66)   &(-0.66, -0.66)     & (-0.66, -0.66) &(-0.66, -0.66)&(-0.66, -0.66)&(-0.65, -0.62)&(-0.70, -0.65)\\ 
\hline
$P_2$	 &  $ 0.72 \pm 0.05$   & (0.71, 0.73)   & (0.68, 0.71)   &  (0.64, 0.69) &  (0.63, 0.67)  &(0.72, 0.73)  &(0.72, 0.73)&(0.73, 0.76)&(0.69, 0.74)\\ 
\hline
$P_3$	 &  $ 0.00 \pm 0.02 $  & (0.00, 0.00)   & (0.00, 0.00)   & (0.00, 0.00)    &(0.00, 0.00)&(0.00, 0.00)&(0.00, 0.00)&(0.00, 0.00)&(0.00, 0.00) \\ 
\hline
$P'_4$	 &  $ -0.64 \pm 0.01 $  & (-0.64, -0.64)  & (-0.64, -0.64) &  (-0.64, -0.64)    & (-0.64, -0.64)& (-0.64, -0.64)& (-0.64, -0.64)& (-0.64, -0.63)& (-0.65, -0.64)\\ 
\hline
$P'_5$	 &  $-1.14 \pm 0.09$   & (-1.16, -1.13)    &  (-1.13, -1.08)   & (-1.10, -1.03)   & (-1.07, -1.01)&(-1.15, -1.14)&(-1.15, -1.15)&(-1.21, -1.16)&(-1.17, -1.09)\\ 
\hline
$P'_6$	 &  $ 0.00 \pm 0.13 $  & (0.00, 0.00)   & (0.00, 0.00)   & (0.00, 0.00)   &(0.00, 0.00) &(0.00, 0.00)&(0.00, 0.00)&(0.00, 0.00)&(0.00, 0.00)\\ 
\hline
$P'_8$	 &  $ 0.00 \pm 0.02 $   & (0.00, 0.00)  &   (0.00, 0.00) & (0.00, 0.00)    &(0.00, 0.00)&(0.00, 0.00)&(0.00, 0.00)&(0.00, 0.00)&(0.00, 0.00)\\ 
 \hline
\end{tabular}}
\caption{Results for the  $q^2$ bin [15-19] $\rm GeV^2$ in the SM as well as for several new physics scenarios for the branching fraction, the longitudinal fraction $f_L$, the $\tau$ forward backward asymmetry $A_{FB}$ and the optimized observables $P_i^{(')}$ in $B \to K^* \tau^+ \tau^-$ decay. Here $B_{\tau\tau}^{K^*} \equiv B(B \to K^* \tau^+ \tau^-)$.}
\label{pred-1}
\end{table*}

The $\tau $ observables include differential branching ratio of $B \to K^* \tau^+ \tau^-$, $f_L$ and   $A_{FB}$. We also consider the optimized angular observables $P_{1,2,3}$ and $P'_{4,5,6,8}$. From these observables, we construct the following $\tau - \mu$ lepton-flavor differences:

\begin{eqnarray}
Q^{\tau \mu}_{B} &=& B(B \to K^* \tau^+ \tau^-) -  B(B \to K^* \mu^+ \mu^-)\,, \nonumber\\
Q^{\tau \mu}_{f_L} &=& f_L(B \to K^* \tau^+ \tau^-) - f_L(B \to K^* \mu^+ \mu^-)\,,\nonumber\\
Q^{\tau \mu}_{A_{FB} }&=& A_{FB} (B \to K^* \tau^+ \tau^-) - A_{FB} (B \to K^* \mu^+ \mu^-)\,, \nonumber\\
Q^{\tau \mu}_{P_i^{(')}} &=& P_i^{(')}(B \to K^* \tau^+ \tau^-) - P_i^{(')}(B \to K^* \mu^+ \mu^-)\,.
\label{defq}
\end{eqnarray}

We also consider the following lepton-flavor ratios:
\begin{eqnarray}
R^{\tau \mu}_{K^*} &=& \frac{ B(B \to K^* \tau^+ \tau^-)}{  B(B \to K^* \mu^+ \mu^-) }\,, \nonumber\\
R^{\tau \mu}_{A_{FB}} &=& \frac{A_{FB}(B \to K^* \tau^+ \tau^-)}{ A_{FB}(B \to K^* \mu^+ \mu^-) }\,,\nonumber\\
R^{\tau \mu}_{f_L}  &=& \frac{f_L (B \to K^* \tau^+ \tau^-)}{ f_L (B \to K^* \mu^+ \mu^-) }\,,\nonumber\\
R^{\tau \mu}_{P_j} &=& \frac{P_j (B \to K^* \tau^+ \tau^-)}{ P_j (B \to K^* \mu^+ \mu^-) }\,, \nonumber\\
R^{\tau \mu}_{P'_k}  &=& \frac{P'_k (B \to K^* \tau^+ \tau^-)}{ P'_k  (B \to K^* \mu^+ \mu^-) }\,,
\label{defr}
\end{eqnarray}
where $j=1,2$ and $k=4,5$. Here we do not consider the following LFU ratios:
\begin{eqnarray}
R^{\tau \mu}_{P_3} &=& \frac{P_3 (B \to K^* \tau^+ \tau^-)}{ P_3 (B \to K^* \mu^+ \mu^-) }\,, \nonumber\\
R^{\tau \mu}_{P'_6} &=& \frac{P'_6 (B \to K^* \tau^+ \tau^-)}{ P'_6   (B \to K^* \mu^+ \mu^-) }\,,\nonumber\\
R^{\tau \mu}_{P'_8} &=& \frac{P'_8 (B \to K^* \tau^+ \tau^-)}{ P'_8   (B \to K^* \mu^+ \mu^-) }\,.
\label{defr1}
\end{eqnarray}
This is due the fact that these observables have large errors due to zero crossings in their $q^2$ spectra.  The lepton-flavor difference and ratio observables have been studied in  $\mu-e$ sector in refs. \cite{Altmannshofer:2014rta,Jager:2014rwa,Altmannshofer:2015mqa,Capdevila:2016ivx,Hurth:2021nsi}.

\begin{figure}[htb!]
\centering
\includegraphics[width = 3.1in]{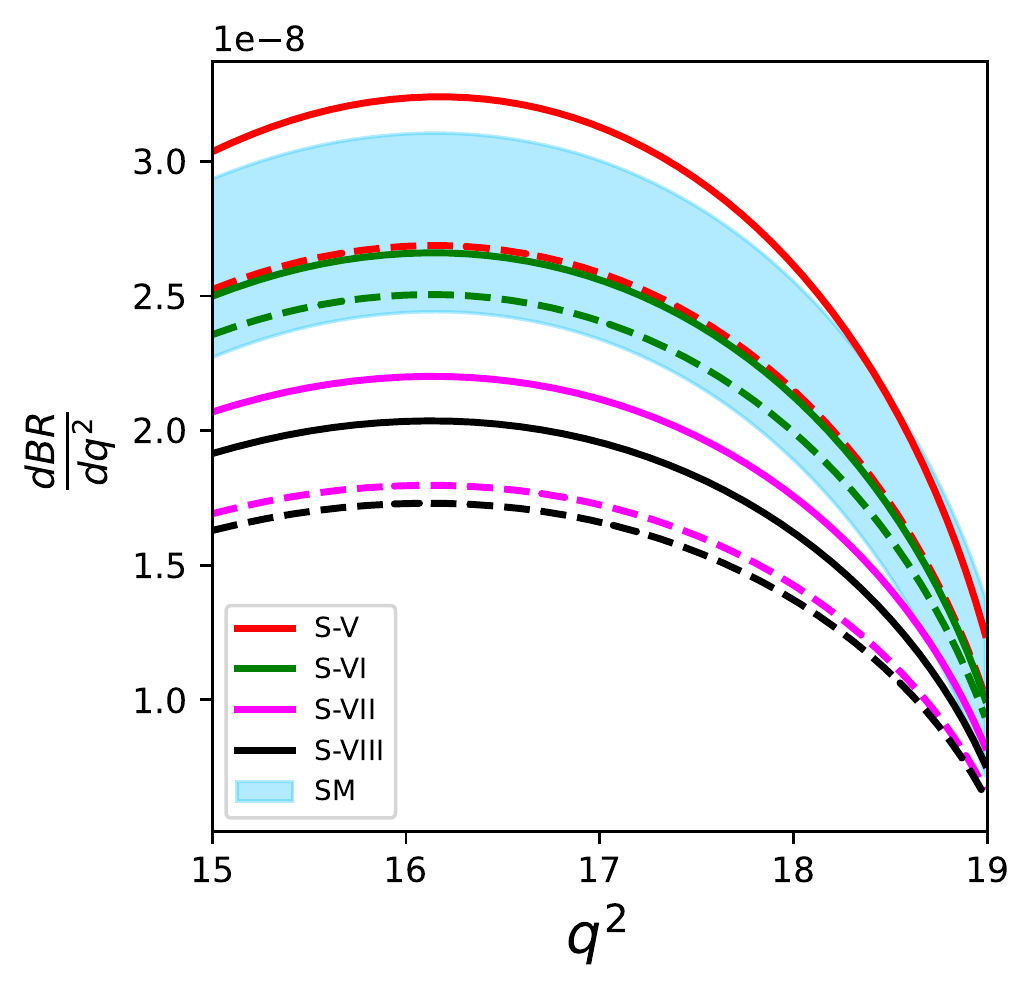}
\caption{Plot for the  $q^2$ distribution in the SM as well as for class-A new physics solutions for the branching fraction of $B \to K^* \tau^+ \tau^-$. The light blue band is due to theoretical uncertainties. The thick and dotted lines represent maximum deviation from the SM for each new physics solutions. We have not shown plots for other $\tau$ observables as their predictions for all considered new physics scenarios only show marginal or negligible deviations from the SM.}
\label{fig:tau1}
\end{figure}

For $B \to K^* \tau^+ \tau^-$ decay, the ditauon $q^2$ ranges from 15 $\rm GeV^2$ to 19.2 $\rm GeV^2$. Within this region, the form factors are computed using a combined fit to lattice and LCSR results.  The predictions of the  branching ratio, $f_L$, $A_{FB}$, optimized angular observables $P_{1,2,3}$ \& $P'_{4,5,6,8}$  in $B \to K^* \tau^+ \tau^-$ decay along with the LFU ratio $R^{\tau \mu}_{K^*}$ are obtained using {\tt flavio} \cite{Straub:2018kue}. We obtained the prediction of other LFU ratios defined above along with the difference observables $Q^{\tau \mu}$  by implementing  them in  {\tt flavio} using the corresponding predefined $\tau$ and $\mu$ observables.

\begin{table*}[htb]
\centering
\resizebox{\textwidth}{!}{ 
\begin{tabular}{|c|c||c|c|c|c||c|c|c|c|c|c|} \hline
Observable & SM  & {S-V}  & {S-VI} & {S-VII} & {S-VIII}& {S-IX}  & {S-X} & {S-XI} & {S-XIII} \\
\hline 
$Q^{\tau \mu}_{f_L}$	&  -0.24 $\pm$ 0.02&    (-0.25, -0.23) &  (-0.23, -0.22) &  (-0.23, -0.22)    &(-0.22, -0.21) &(-0.24, -0.24) &(-0.24, -0.24) &(-0.24, -0.24) &(-0.24, -0.23)  \\ 
\hline
$Q^{\tau \mu}_{A_{FB}}$	& -0.15 $\pm$ 0.02 &   (-0.19, -0.06) &  (-0.14, -0.11)   &   (-0.12, -0.06)   &(-0.11, -0.08)  &(-0.16, -0.15)&(-0.16, -0.13)&(-0.15, -0.13)&(-0.20, -0.12)\\ 
\hline
$Q^{\tau \mu}_{P_1}$	 & -0.037$\pm$ 0.002 &   (-0.04, -0.04)  &  (-0.04, -0.04) &  (-0.04, -0.04) & (-0.04, -0.04)  &(-0.04, -0.03) &(-0.04, -0.04) &(-0.05, -0.04) & (-0.14, -0.06)   \\  
\hline
$Q^{\tau \mu}_{P_2}$	 &0.35$\pm$ 0.02 &     (0.35, 0.43)  &  (0.33, 0.35) &(0.32, 0.35) &(0.31, 0.33) &(0.35, 0.35)&(0.36, 0.37)&(0.37, 0.39)&(0.32, 0.38)\\ 
\hline
$Q^{\tau \mu}_{P_3}$	 & 0.000 $\pm$ 0.001&    (0.00, 0.00) &  (0.00, 0.00)   & (0.00, 0.00)    &   (0.00, 0.00)&(0.00, 0.00)&(0.00, 0.00)&(0.00, 0.00)&(-0.01, 0.00)\\  
\hline
$Q^{\tau \mu}_{P'_4}$	 & -0.008 $\pm$ 0.001 &     (-0.01, -0.01)  &  (-0.01, -0.01)  &  (-0.01, -0.01)  &   (-0.01, -0.01) &(-0.01, -0.01) &(-0.01, -0.01) &(-0.01, -0.01) &(-0.03, -0.01) \\ 
\hline
$Q^{\tau \mu}_{P'_5}$	 & -0.55 $\pm$  0.06&   (-0.67, -0.54) &  ( -0.55,  -0.53)&(-0.56, -0.50)    &(-0.52, -0.49) &(-0.56, -0.55) &(-0.58, -0.56) &(-0.62, -0.58) &(-0.58, -0.47) \\ 
\hline
$Q^{\tau \mu}_{P'_6}$	 & 0.00 $\pm$ 0.06 &    (0.00, 0.00) &  (0.00, 0.00) &  (0.00, 0.00)  &   (0.00, 0.00)&(0.00, 0.00)&(0.00, 0.00)&(0.00, 0.00)&(0.00, 0.00)\\
\hline
$Q^{\tau \mu}_{P'_8}$	 & 0.000 $\pm$ 0.001 &   (0.00, 0.00)   & (0.00, 0.00)  &  (0.00, 0.00)     & (0.00, 0.00)&(0.00, 0.00)&(0.00, 0.00)&(0.00, 0.00)&(0.00, 0.01)\\ 
 \hline
\end{tabular}}
\caption{Results for the  $q^2$ bin [15-19] $\rm GeV^2$ in the SM as well as for several new physics scenarios for the $\tau-\mu$ lepton-flavor difference observables (the $Q^{\tau \mu}$ observables) for the longitudinal fraction $f_L$, the forward-backward asymmetry $A_{FB}$  and the optimized observables  $P_i^{(')}$. }
\label{pred-2}
\end{table*} 

In the following, we provide integrated values of all considered observables  in [15-19] $q^2$ bin, which is the only bin relevant for $B \to K^* \tau^+ \tau^-$ decay. The $q^2$ graphs will be shown only for those observables for which a noticeable deviation from the SM, say more than 25\%, is allowed for atleast one of the new physics solutions. Further, we  perform separate analysis for class-A and B solutions, i.e., we will firstly compare amongst various allowed solutions within one class and then look for possible distinction between the two classes of new physics solutions. 

The predictions of integrated values of $\tau$ observables in [15-19] $q^2$ bin are exhibited in Tab.~\ref{pred-1}.   
It is evident that  the predictions for the branching ratio of $B \to K^* \tau^+ \tau^-$ for scenarios V and VI are consistent with the SM whereas the S-VII and S-VIII solutions can lead to a suppression of up to $\sim$ 25\% in the value of $B(B \to K^* \tau^+ \tau^-)$. This is also reflected from the graph of differential branching ratio which is portrayed in Fig.~\ref{fig:tau1}. It can be seen that the 1$\sigma$ allowed region for S-VII and  S-VIII do not overlap with the corresponding SM range. Further, no notable deviation is allowed for any of the class-B new physics solutions i.e., solutions  characterized either by $C_{10}^U$ or $C_{10}^{\prime U}$  contributions.

The allowed values of longitudinal polarization fraction $f_L$ for all solutions are consistent with the SM predictions. This includes class-B solutions as well. The same is true for $A_{FB}$ and $P_1$.  The angular observable $P_2$ can be  suppressed by $\sim$ 5\% as compared to the SM for the new physics solutions S-VII and S-VIII. The remaining  scenarios do not provide any interesting features.

The value of $P_3$ in SM is $0.00\pm 0.02$. As illustrated in the Tab.~\ref{pred-1}, none of the new physics scenarios can provide any useful enhancement in the $P_3$ observable. The value of $P'_4$ in the SM is $\sim -0.6$ which remains almost the same for all considered scenarios. The SM prediction for  $P'_5$  observable   is $\sim -1$. Any large deviation in this observable is forbidden by the current $b \to s \ell \ell $ data as can be seen from  Tab.~ \ref{pred-1}. Within the SM, the value of observable $P'_6$ is $0.00 \pm 0.13$.  All  four scenarios considered in our analysis do not show any improvement over SM value as culminated from the table.
For $P'_8$, the results are in the similar lines to that of observable $P_3$.

Thus it is evident from Tab.~\ref{pred-1} that none of the new physics solutions considered in this work can provide a large deviation, say 50\% or more, from the SM prediction in any of the $\tau$-observables under consideration. Only a  deviation of 25\% is allowed in the branching ratio of $B \to K^* \tau \tau$ for S-VII and S-VIII solutions which belong to the class-A scenario.   It should be noted that this deviation is in the form of suppression, i.e., the current new physics solutions can only provide suppression in the $\tau$ observables. As far as discriminating new physics solutions are concerned, the fact that deviation is not  substantial, a very precise measurement  would be required. None of the class-B solutions can provide any notable deviation from the SM in any of the $\tau$ observables. Therefore any observational suppression in $B(B \to K^* \tau^+ \tau^-)$ would provide evidence in support of class-A new physics in the form of the S-VII and S-VIII solutions.

We now consider LFU observables constructed by taking difference of observables in the decay of $B \to K^* \tau^+ \tau^-$ and $B \to K^* \mu^+ \mu^-$. These $Q^{\tau \mu}$ observables are defined in Eqs.~\eqref{defq}. The prediction of $Q^{\tau \mu}$ observables for the  $q^2$ bin [15-19] $\rm GeV^2$ in the SM as well as for considered new physics scenarios are shown in Tab.~\ref{pred-2}.

\begin{figure}[htb]
\centering
\includegraphics[width = 3.1in]{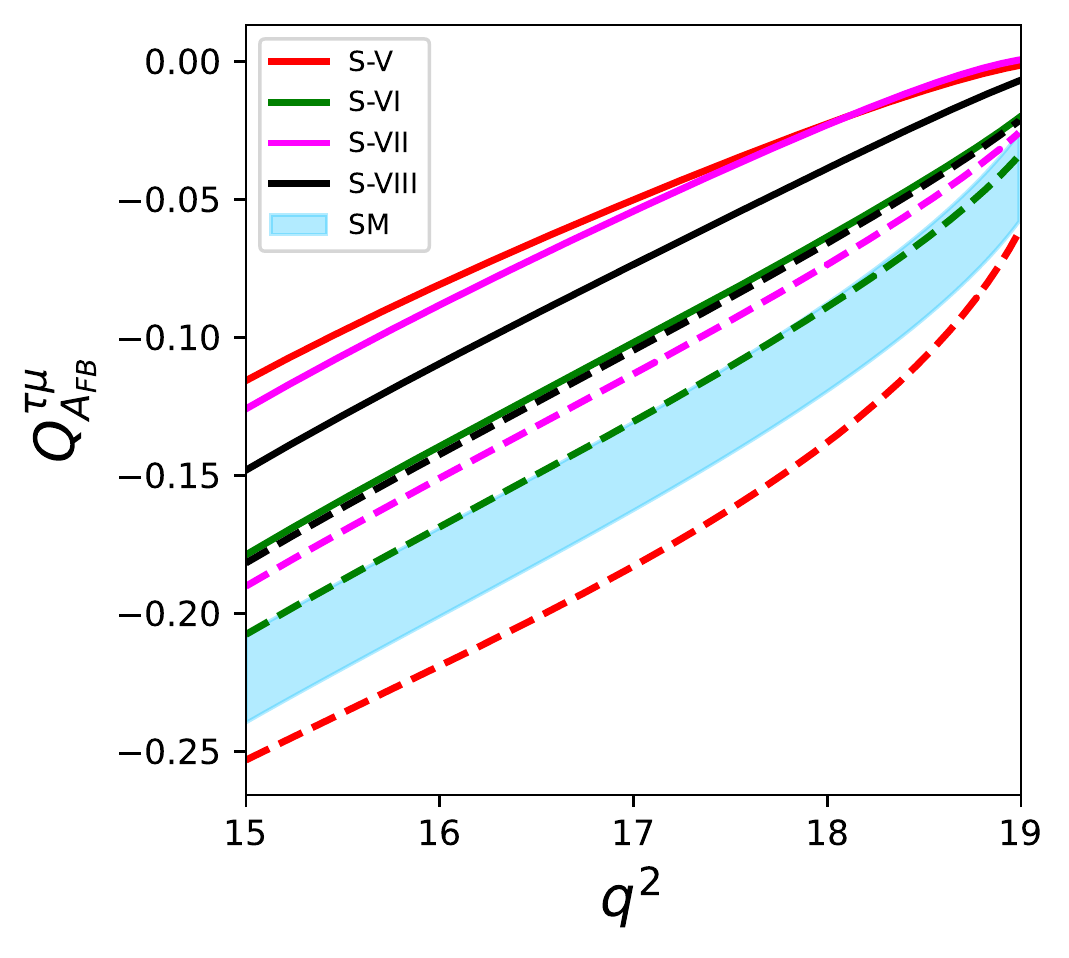}
\caption{Plot for the  $q^2$ distribution in the SM as well as for several class-A new physics solutions for the $\tau-\mu$ lepton-flavor difference observable, the $Q^{\tau \mu}$ observable,  for the forward backward asymmetry $A_{FB}$. The light blue band is due to theoretical uncertainties. The thick and dotted lines represent maximum deviation from the SM for each new physics solutions.}
\label{fig:q1}
\end{figure}

\begin{figure*}[htb]
\centering
\includegraphics[width = 3.1in]{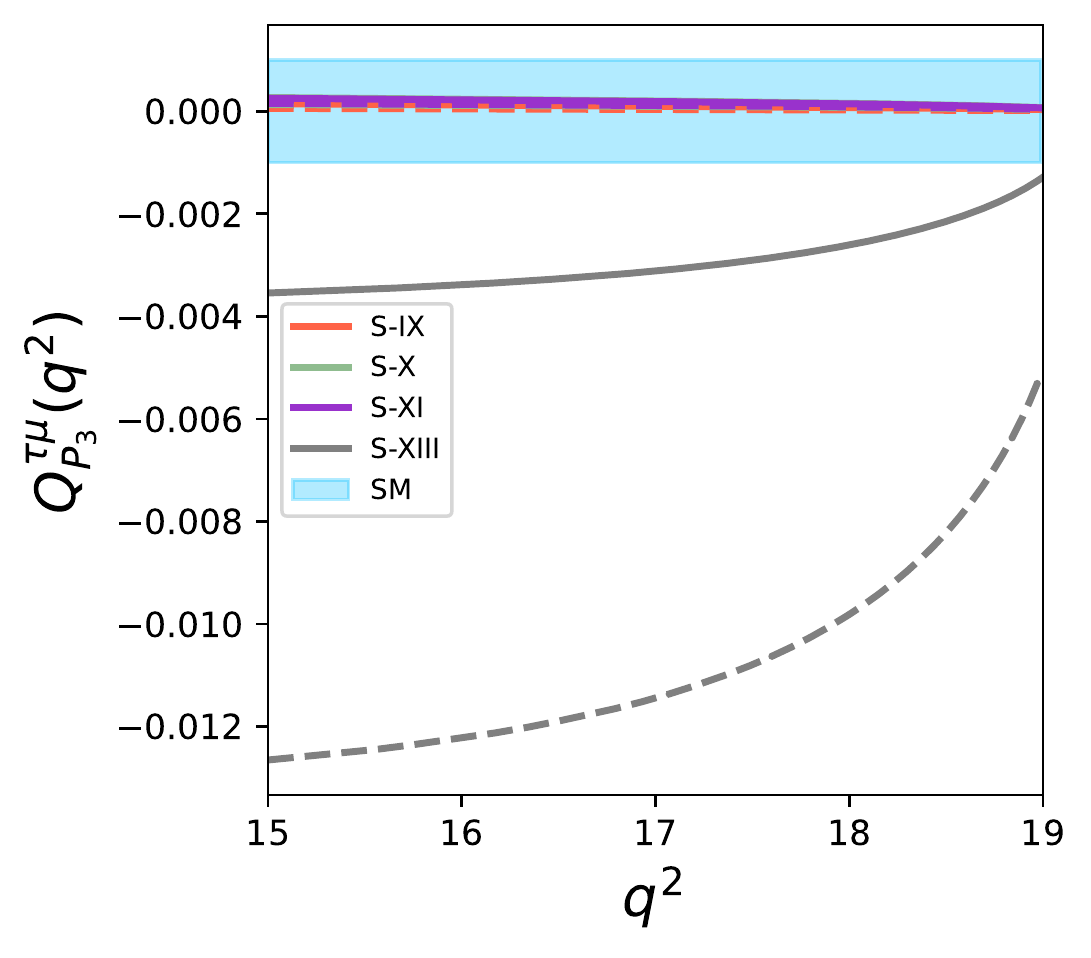}
\includegraphics[width = 3.05 in]{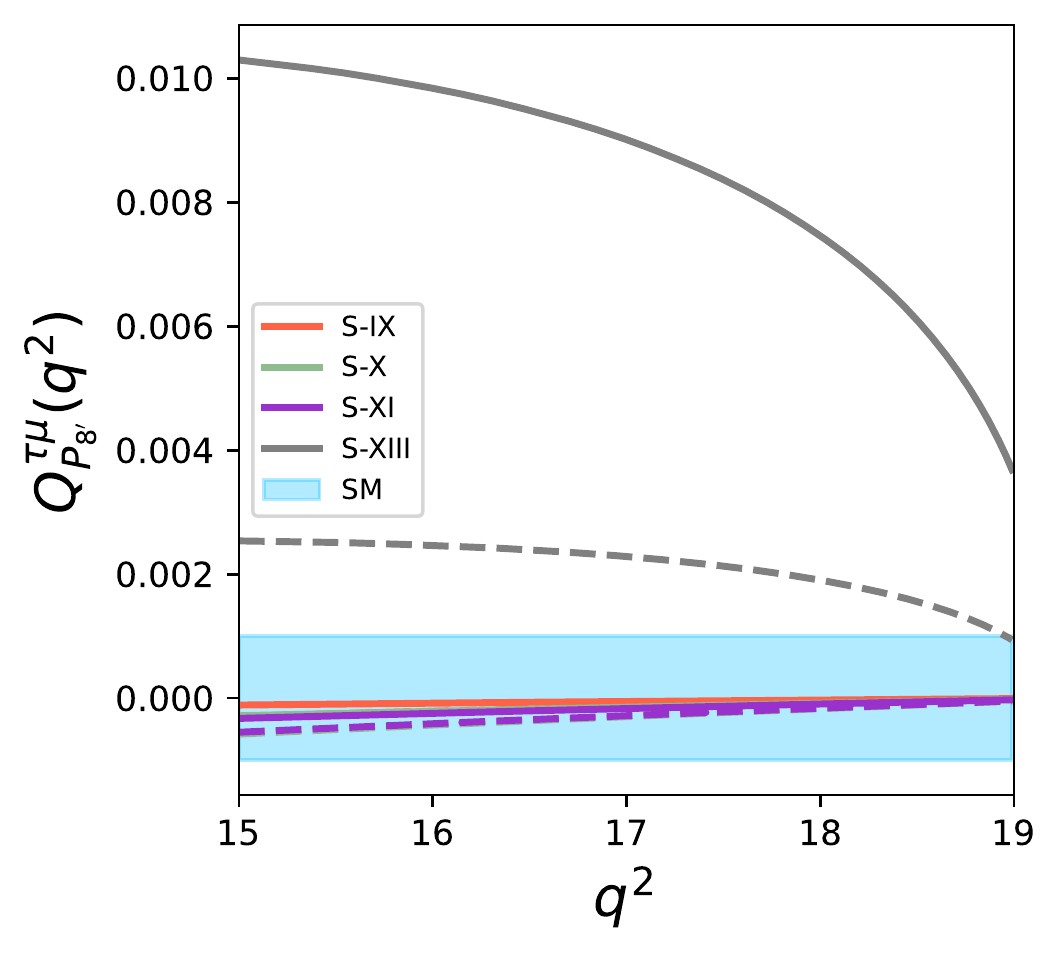}
\caption{Plots for the  $q^2$ distribution in the SM as well as for class-B new physics scenarios for the $\tau-\mu$ lepton-flavor difference observables, the $Q^{\tau \mu}$ observables,  for optimized observables $P_{3}$ and  $P'_{8}$. The light blue band is due to theoretical uncertainties. The thick and dotted lines represent maximum deviation from the SM for each new physics solutions. }
\label{fig:tq1}
\end{figure*}

The LFU difference between the longitudinal polarization fractions, $Q^{\tau \mu}_{f_L}$, is predicted to be $\sim -0.25$ in the SM. This predicted value remains the same for all new physics scenarios under consideration, including class-B solutions. The observable $Q^{\tau \mu}_{A_{FB}}$ provides promising features as its value can be enhanced as compared to the SM for S-V and S-VII solutions. These solutions can provide an amelioration  up to twofold in  the value of $Q^{\tau \mu}_{A_{FB}}$  above its SM prediction.  
The same feature is also reflected in the $q^2$ distribution plot of $Q^{\tau \mu}_{A_{FB}}$ as shown in Fig.~\ref{fig:q1}. It should be noted that none of the class-B solutions can provide any visible deviation from the SM prediction of $Q^{\tau \mu}_{A_{FB}}$. Thus any prominent deviation in this observable would disfavor class-B solutions.

We now analyze lepton-flavor differences for optimized observables. We firstly consider class-A solutions. 
The prediction  of $Q^{\tau \mu}_{P_1}$ observable for all class-A solutions are consistent with the SM. The SM value of observable $Q^{\tau \mu}_{P_2}$ is $\sim 0.35$. An enhancement of 10\% is possible due to scenario S-V whereas $Q^{\tau \mu}_{P_2}$ for other class-A solutions are consistent with the SM prediction. The SM prediction of $Q^{\tau \mu}_{P_3}$ is negligibly small, $\sim 10^{-3}$. Therefore this observable can be measured only if any new physics contributions can provide a very large enhancement. However, as can be seen from Tab.~\ref{pred-2}, none of the class-A new physics scenarios can effectuate any meaningful enhancement in the value of $Q^{\tau \mu}_{P_3}$. The same is true for the $Q^{\tau \mu}_{P'_8}$  observable.

\begin{table*}[t]
\centering
\resizebox{\textwidth}{!}{ 
\begin{tabular}{|c|c||c|c|c|c||c|c|c|c|c|c|} \hline
Observable & SM  & {S-V}  & {S-VI} & {S-VII} & {S-VIII}& {S-IX}  & {S-X} & {S-XI} & {S-XIII} \\ 
\hline 
$R^{\tau \mu}_{K^*}$	   & $0.41 \pm 0.01$&  (0.37, 0.67)   &   (0.39, 0.46)   &(0.34, 0.40) &  (0.34, 0.39) &(0.43 0.49) &(0.45, 0.51)&(0.46, 0.51)&(0.48, 0.60)\\ 
\hline
$R^{\tau \mu}_{f_L}$	&  0.30 $\pm$ 0.01&   (0.25, 0.31)   &   (0.31, 0.34)  &  (0.33, 0.36)  &  (0.34, 0.37)&(0.28, 0.30)&(0.28, 0.29)&(0.29, 0.29)& (0.28, 0.31)\\ 
\hline
$R^{\tau \mu}_{A_{FB}}$	& 0.58 $\pm$ 0.02&   (0.49, 0.77)  &   (0.61, 0.67) &  (0.65, 0.79) & (0.67, 0.74)&(0.56, 0.59)&(0.57, 0.61)&(0.59, 0.62)& (0.51, 0.62) \\ 
\hline
$R^{\tau \mu}_{P_1}$	 & 1.05 $\pm$ 0.01 &  (1.06, 1.06)  &  (1.06, 1.06) &  (1.06, 1.06)   & (1.06, 1.06)&(1.06, 1.06)&(1.06, 1.06)&(1.06, 1.09)&(1.09, 1.28)\\  
\hline
$R^{\tau \mu}_{P_2}$	&1.94 $\pm$ 0.01 &  (1.92, 2.52)    & (1.95, 1.97) &  (1.96, 2.15)  & (1.94, 1.98)&(1.94, 1.95)&(1.97, 2.06)&(2.01, 2.09)& (1.83, 2.06)\\ 
\hline
$R^{\tau \mu}_{P'_4}$	 &1.012 $\pm$ 0.001  & (1.01, 1.01)     & (1.01, 1.01)   &(1.01,  1.01)  & (1.01, 1.01)&(1.01,  1.01)&(1.01,  1.01)&(1.01,  1.02)&(1.02,  1.05) \\ 
\hline
$R^{\tau \mu}_{P'_5}$	 & 1.93 $\pm$ 0.01&   (1.91, 2.47)   & (1.94, 1.96) &  (1.95, 2.13) & (1.93, 1.96)&(1.93, 1.94)&(1.96, 2.03)&(1.98, 2.06)&(1.76, 2.02)\\    \hline
\end{tabular}}
\caption{Results for the  $q^2$ bin [15-19] $\rm GeV^2$ in the SM as well as for several new physics scenarios for the $\tau-\mu$ lepton-flavor  ratio observables, the $R^{\tau \mu}$ observables,   the longitudinal fraction $f_L$, for the forward-backward asymmetry $A_{FB}$, and several angular observables  $P_i^{(')}$. Here we do not consider $R^{\tau \mu}_{P_3}$, $R^{\tau \mu}_{P'_6}$ and $R^{\tau \mu}_{P'_8}$ ratios as they have large errors due to zero crossing.}
\label{pred-3}
\end{table*}

\begin{figure*}[htb]
\centering
\includegraphics[width = 3.1in]{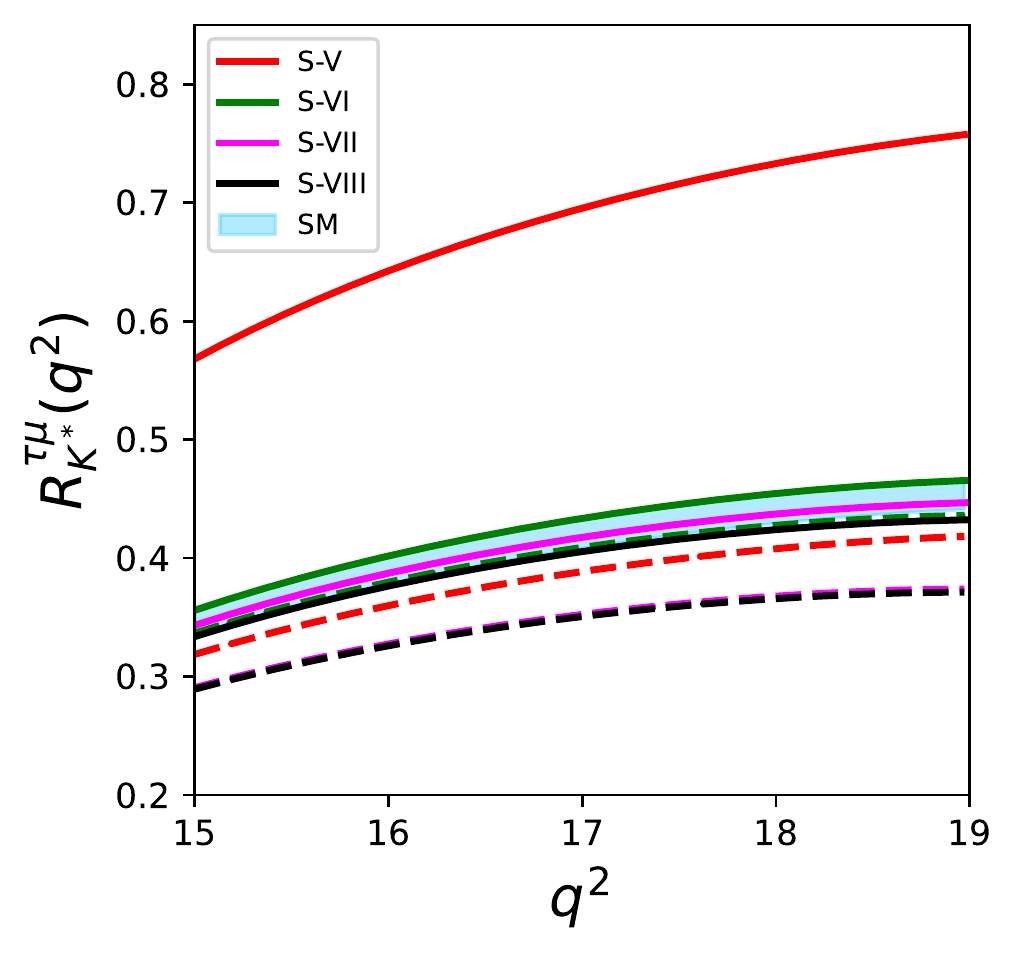}
\includegraphics[width = 3.1in]{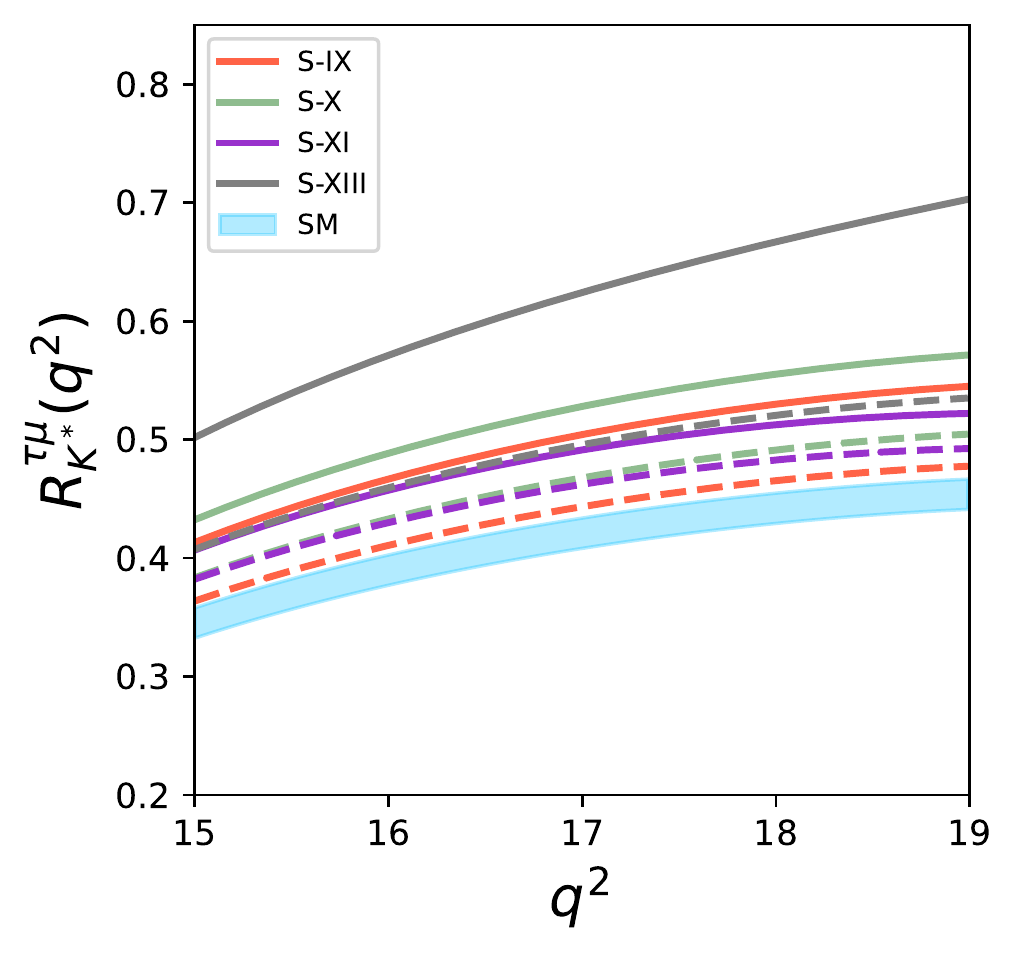}
\caption{Plots for the  $q^2$ distribution in the SM as well as for several new physics solutions for the $\tau-\mu$ LFU ratio, the $R^{\tau \mu}$ observable,  for the  branching fraction. The left and right panels correspond to the predictions for class-A and class-B solutions, respectively.  The light blue band is due to theoretical uncertainties. The thick and dotted lines represent maximum deviation from the SM for each new physics solutions.}
\label{fig:r1}
\end{figure*}

\begin{figure*}[htb]
\centering
\includegraphics[width = 3.1in]{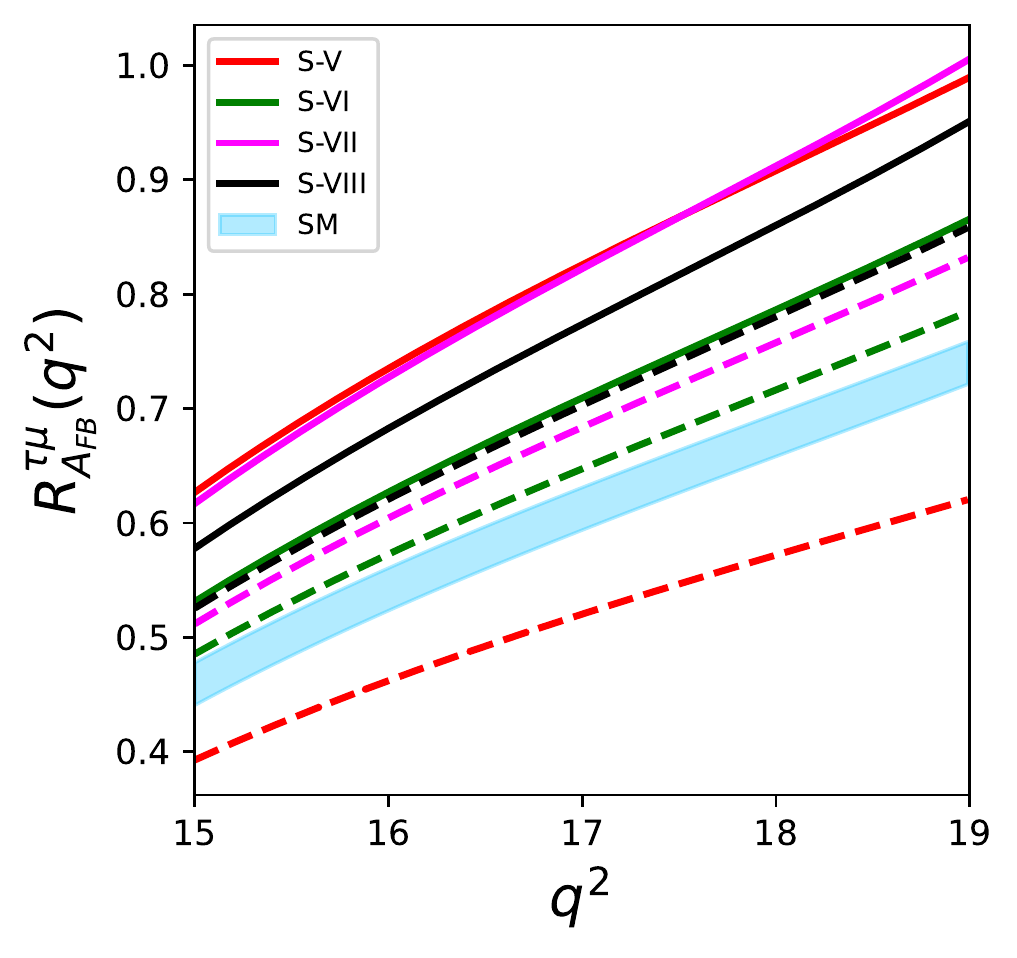}
\includegraphics[width = 3.1in]{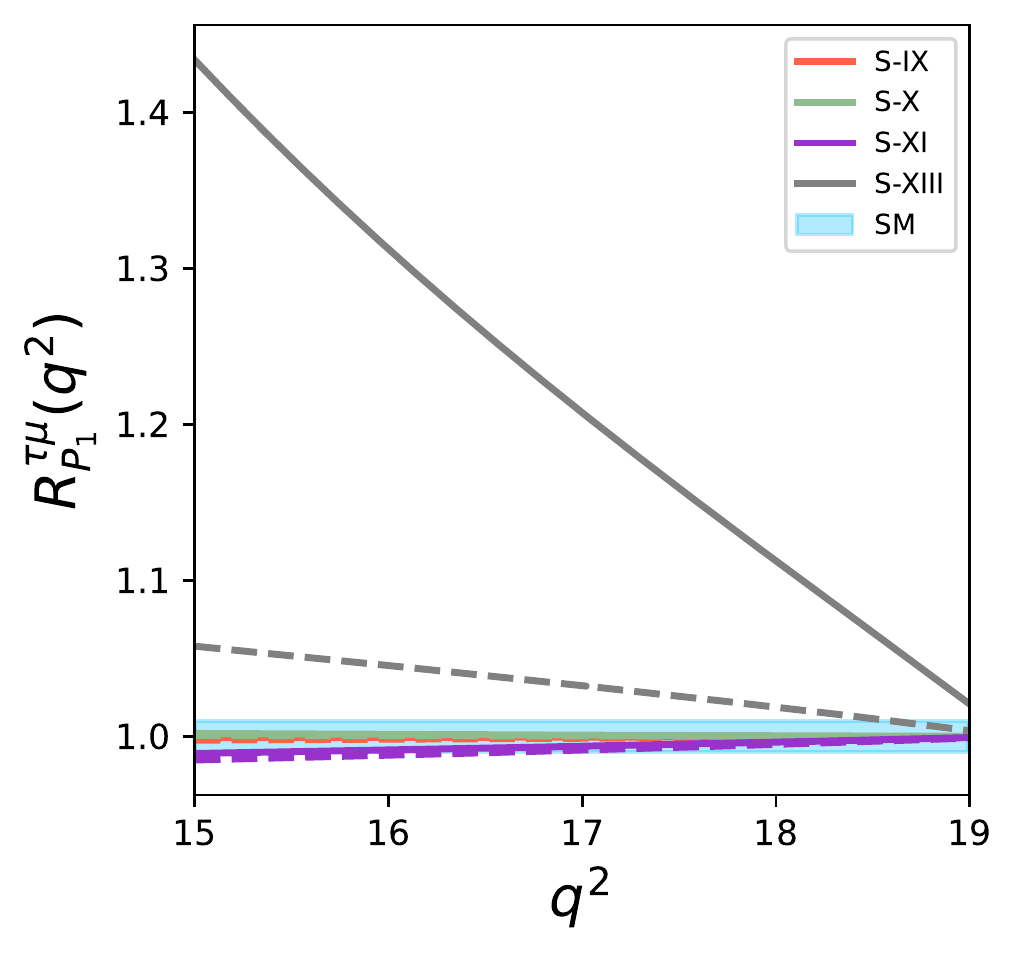}
\caption{Plots for the  $q^2$ distribution in the SM as well as for several new physics solutions for the $\tau-\mu$ LFU ratio, the $R^{\tau \mu}$ observables. The left and right panels correspond to the predictions for $R^{\tau \mu}_{A_{FB}}$ (class-A solutions) and $R^{\tau \mu}_{P_1}$ (class-B solutions), respectively. The light blue band is due to theoretical uncertainties. The thick and dotted lines represent maximum deviation from the SM for each new physics solutions.}
\label{fig:rafb}
\end{figure*}

\begin{figure*}[htb]
\centering
\includegraphics[width = 3.1in]{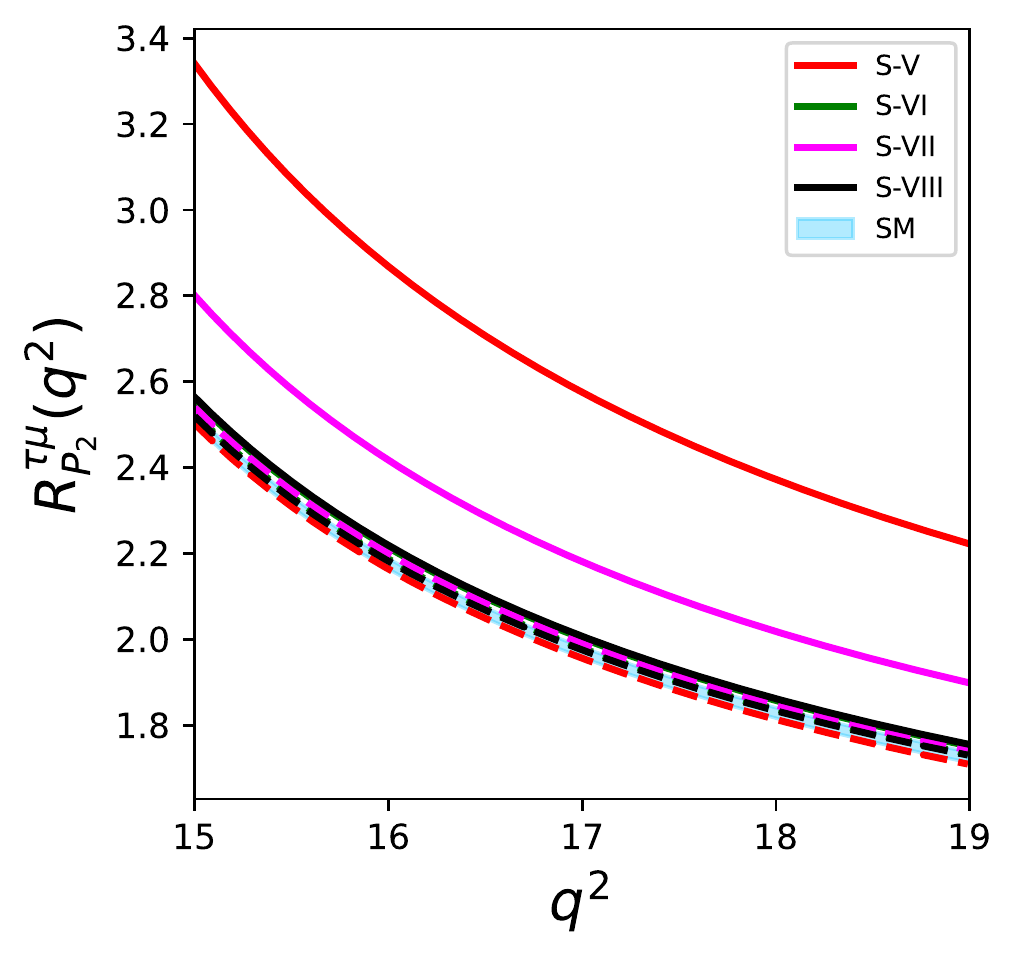}
\includegraphics[width = 3.1in]{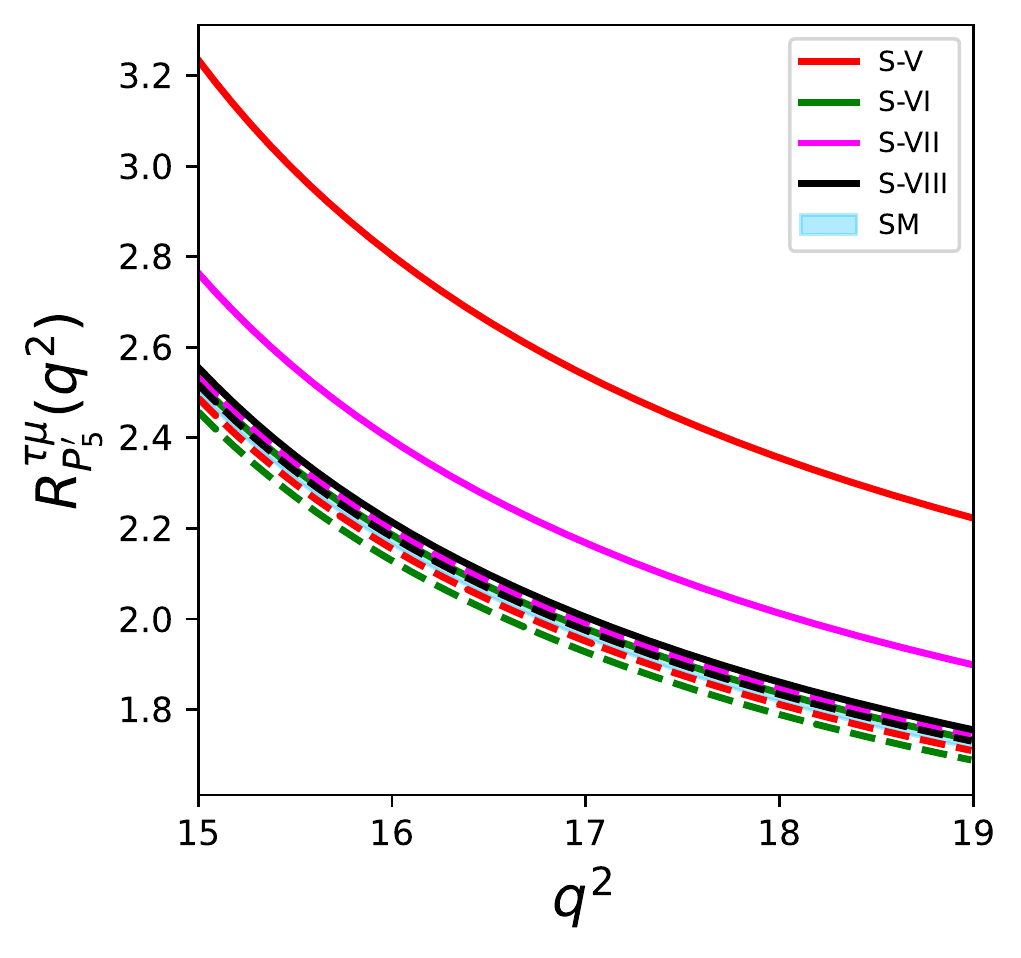}
\caption{Plots for the  $q^2$ distribution in the SM as well as for several class-A new physics scenarios for the $\tau-\mu$ LFU ratios, the $R^{\tau \mu}$ observables,  for the optimized observables $P_2$ (left panel) and  $P'_5$ (right panel). The light blue band is due to theoretical uncertainties. The thick and dotted lines represent maximum deviation from the SM for each new physics solutions.}
\label{fig:r2}
\end{figure*}

The SM prediction of $Q^{\tau \mu}_{P'_4}$ is $\sim -0.01$. None of the class-A new physics  solutions can generate any useful deviation from the SM. The observable $Q^{\tau \mu}_{P'_5}$ is predicted to be $\sim $ -0.5 in the SM.  None of the class-A scenarios show interesting results for $Q^{\tau \mu}_{P'_5}$  observable. The SM prediction of $Q^{\tau \mu}_{P'_6}$ is  $0.00 \pm 0.06$ which remains almost the same for all class-A new physics solutions. Thus we see that none of the class-A solutions can provide any noticeable deviation from the SM in any of the  $Q^{\tau \mu}_{P^{(')}_i}$ observables.

The situation, however, seems to be encouraging for class-B solutions, in particular the S-XIII solution. As can be seen from Tab.~\ref{pred-2}, the S-XIII solution can provide large deviations from the SM in a number of $Q^{\tau \mu}_{P^{(')}_i}$ observables.  For $Q^{\tau \mu}_{P_1}$ observable, a threefold deviation from SM is possible for S-XIII solution.  
For $Q^{\tau \mu}_{P_2}$ observable, all solutions 
cannot dispense any illustrious difference.
A tenfold enhancement in the magnitude of $Q^{\tau \mu}_{P_3}$ is allowed for the S-XIII solution. All other solutions predict SM-like scenario for this observable.

The $Q^{\tau \mu}_{P'_{5,6}}$ observables are predicted to be close to their SM values for all class-B solutions.  A threefold enhancement in the magnitude of $Q^{\tau \mu}_{P'_4}$ is allowed for the S-XIII solution. The deviation is negligible for other solutions. The S-XIII solution can effectuate an order of magnitude enhancement in the magnitude of $Q^{\tau \mu}_{P'_{8}}$ observable. Similar features are also delineated in the $q^2$ graphs, Fig.~\ref{fig:tq1}, of these $Q^{\tau \mu}$ observables. Thus we see that $Q^{\tau \mu}_{P^{(')}_i}$ observables can be useful in discriminating between the class-A and class-B scenarios. Any propitious
deviation in these observables can only be due to class-B solutions, particularly the S-XIII solution.

Finally, we consider $\tau-\mu$ lepton-flavor ratios  of the  branching fractions, the longitudinal fractions, the forward backward asymmetries and the $P_i^{(')}$ angular observables in $B \to K^* \ell \ell$ decay. The prediction for these observables in the SM as well as for favored new physics solutions belonging to class-A as well as class-B scenarios are given in  Tab.~\ref{pred-3}. 

The SM prediction of $R^{\tau \mu}_{K^*}$ is $\sim$ 0.4.  It is apparent from Tab.~\ref{pred-3} that amongst the class-A new physics solutions, the S-V solution can engender largest enhancement in  $R^{\tau \mu}_{K^*}$, by 60\%, from the SM.     The scenarios S-VII and S-VIII can suppress the value of $R^{\tau \mu}_{K^*}$, $\sim$15\% below the SM. These features are also articulated in the $q^2$ distribution of  $R^{\tau \mu}_{K^*}$ which is depicted in the left panel of Fig.~\ref{fig:r1}. 

The class-B solutions can also relinquish large enhancements from the SM value. The largest enhancement is allowed for the S-XIII solution. This can enhance $R^{\tau \mu}_{K^*}$  by $\sim$40\%. The S-IX, S-X and S-XI solutions can also induce enhancement in $R^{\tau \mu}_{K^*}$. However these enhancements cannot exceed by more than $\sim$20\%. Further, it can be espied from the right panel of Fig.~\ref{fig:r1} that none of the class-B solutions can lead to suppression in  $R^{\tau \mu}_{K^*}$  below the SM value.  

The SM prediction of $\tau-\mu$ flavor ratio of the longitudinal polarization asymmetries,  $R^{\tau \mu}_{f_L}$, is $\sim 0.3$.  None of the class-A new physics scenarios, except S-V,  can provide any meaningful suppression in its value. Even for S-V, the suppression can only be up to $\sim$15\%  The new physics solution S-VII and S-VIII can ameliorate  $R^{\tau \mu}_{f_L}$ by $\sim$20\% above the SM.  Further, none of the class-B solutions can invoke any divergence from the SM value of $R^{\tau \mu}_{f_L}$. 

Within the SM, the $\tau-\mu$ flavor ratio of the forward backward asymmetries,  $R^{\tau \mu}_{A_{FB}}$,  attains a value of $\sim 0.6$. It should be noted that this ratio for $\mu-e$ observable, $R^{\mu, e}_{A_{FB}}$, in the [1, 6] $\rm GeV^2$ region is not a good observable to probe new physics due to large errors in its SM prediction. This is due to the fact that the the $q^2$ distribution of $A_{FB}$ exhibits a zero crossing in this bin. For $R^{\tau \mu}_{A_{FB}}$, the relevant $q^2$ bin is  [15, 19] $\rm GeV^2$ for which there is no zero-crossing.

 Amongst all class-A solutions, the S-V and S-VII solutions can provide largest enhancement in $R^{\tau \mu}_{A_{FB}}$. This is apparent from the left panel of Fig.~\ref{fig:rafb}. The enhancement can be up to 30\%. None of the class-A new physics solutions can deplete  $R^{\tau \mu}_{A_{FB}}$, except S-V which can induce $\sim$15\% depletion.  Further, none of the class-B solutions can render large deviations from the SM.

We now consider $\tau-\mu$ flavor ratio of the optimized angular observables in $B \to K^* \ell \ell$ decay. The flavor ratio $R^{\tau \mu}_{P_1}$ is predicted to be $\sim 1$ in the SM. None of the class-A new physics solutions can provide any useful deviation from the SM. The same is true for the $R^{\tau \mu}_{P'_4}$ observable for which the SM prediction is close to unity and all class-A new physics scenarios fail to provide any impact. The status remains the same for the $R^{\tau \mu}_{P'_4}$ observable for class-B solutions, i.e. none of the class-B solutions can provide any visible deviation from the SM. However, as seen from the right panel of Fig.~\ref{fig:rafb}, for  $R^{\tau \mu}_{P_1}$ observable, the S-XIII solution can invoke $\sim$ 20\% boost over the SM value. 

The flavor ratio  $R^{\tau \mu}_{P_2}$ is predicted to be $\sim 2$ in the SM. For class-A scenario, the S-V scenario can provide 30\% enhancement over the SM value.  Apart from  
S-V, the S-VII scenario can also enhance $R^{\tau \mu}_{P_2}$ above the SM value, the enhancement can only be up to $\sim 10\%$. These features are reflected in the $q^2$ plot of $R^{\tau \mu}_{P_2}$, the left panel of Fig.~\ref{fig:r2}. None of the class-B solutions can generate large new physics effects in $R^{\tau \mu}_{P_2}$ observable.  This is evident from Tab.~\ref{pred-3}.
 The results are almost the same for the $R^{\tau \mu}_{P'_5}$ observable. Within the SM, this observable is predicted to be $\sim 2$. For class-A solutions, the largest enhancement from the SM  is provided by the S-V solution, $\sim 25\%$, as can be seen from the right panel of Fig.~\ref{fig:r2}. The scenario S-VII can provide enhancement but only up to $\sim$10\%. The class-B solutions fail to generate any observable impact on the $R^{\tau \mu}_{P'_5}$ observable except S-XIII solution which can induce marginal depletion in $R^{\tau \mu}_{P'_5}$, up to $\sim 10\%$.
 
 Thus we see that the current data in $b \to s \mu^+ \mu^-$  sector allows for large deviations in a number of observables related to the decay of $B \to K^* \tau^+ \tau^-$. These observables will be particularly interesting in  hunting for violation of lepton-flavor universality in $\mu-\tau$ sector.
 However, the situation is not so encouraging from the experimental front due to presence of multiple neutrinos in the final state. Therefore, in order to utilize the potential of $B \to K^* \tau^+ \tau^-$ decay, a significant improvement in the current $\tau$ reconstruction techniques would be required.

%%%%%%%%%%%%%%%%%
 \section{Conclusions}
\label{concl}
%%%%%%%%%%%%%%%%%%%%%%%%%%%
We study the impact of  current $b \to s \ell^+ \ell^-$ ($\ell=e,\,\mu$) measurements on several observables in the decay of $B \to K^* \tau^+ \tau^-$  under the assumption that the possible new physics contributions to $b \to s \ell^+ \ell^-$ can have both universal as well as  nonuniversal couplings to leptons. The analysis is performed in a model agnostic way using the language of effective field theory. The primary goal is to identify observables where large new physics effects are allowed by  scenarios which provide a good fit to all $b \to s \ell \ell$ data. We also intend to discriminate between various new physics solutions which are classified in two categories: solutions having universal $C_9$ couplings   and solutions with universal $C_{10}$ or $C_{10}^{\prime}$  couplings. We denote them as class-A and class-B scenarios, respectively. In our analysis, we consider a number of observables related  to $B \to K^* \tau^+ \tau^-$. These include the branching fraction, the $K^*$ longitudinal fraction $f_L$, the tau forward backward asymmetry $A_{FB}$ as well as optimized angular  observables  $P_{1,2,3}$ and  $P'_{4,5,6,8}$. We then construct $\tau - \mu$  LFUV difference (ratio) observables by taking differences (ratios) of  branching fractions, $f_L$'s, $A_{FB}$'s and $P_i^{(')}$'s of $B \to K^* \tau^+ \tau^-$ and $B \to K^* \mu^+ \mu^-$ decays. 

For $\tau$ observables, i.e., for observables related only to $B \to K^* \tau^+ \tau^-$ decay, we observe the following:
\begin{itemize}
\item None of the allowed solutions can generate notable enhancement in the branching fraction of $B \to K^* \tau^+ \tau^-$ decay. In fact,  new physics solutions S-VII $\equiv \left(C^{V}_{9\mu}, \, C^{U}_9\right)$ and S-VIII $\equiv \left(C^{V}_{9\mu} = - C^{V}_{10\mu},\, C^{U}_9 \right)$ belonging to class-A category can  induce suppression up to $\sim 25\%$ as compared to the SM. No notable depletion in branching fraction is possible for any of the class-B new physics solutions.  

\item The $K^*$ longitudinal fraction $f_L$ and the tau forward backward asymmetry $A_{FB}$ are predicted to be close to their SM values for all allowed solutions.

\item No noticeable new physics effects are allowed in any of the optimized angular observables for all new physics solutions.
\end{itemize}

The results for $\tau - \mu$  LFUV difference observables, $Q^{\tau \mu}$, can be summarized as follows:
\begin{itemize}

\item A two fold enhancement in  $Q^{\tau \mu}_{A_{FB}}$ is allowed for S-V  $\equiv ( C^{V}_{9\mu} ,\, C^{V}_{10\mu},\, C^{U}_9 = C^{U}_{10})$  and S-VII solutions. None of the class-B solutions can induce any meaningful enhancement.  

\item A new physics solution, S-XIII $\equiv (C_{9\mu}^V,\, C_{9\mu}^{\prime V},\, C_{10}^{U},\, C_{10}^{\prime U})$ belonging to the class-B scenario can provide  an order of magnitude enhancement  in the absolute values of $Q^{\tau \mu}_{P_3}$ and $Q^{\tau \mu}_{P'_8}$ observables whereas an enhancement up to  threefold is allowed for $Q^{\tau \mu}_{P_1}$ and $Q^{\tau \mu}_{P'_4}$ observables.  None of the class-A solutions can provide any noticeable deviation from the SM in any of the  $Q^{\tau \mu}_{P^{(')}_i}$ observables.
\end{itemize}
 
Finally, we analyze $\tau - \mu$  LFUV ratio observables, $R^{\tau \mu}$.  Our main findings are as follows:
\begin{itemize}
\item The  ratio of branching fractions of $B \to K^* \tau^+ \tau^-$ and
$B \to K^* \mu^+ \mu^-$  can be can be enhanced up to 40\% - 50\% over the SM value. This enhancement is possible for S-V as well as S-XIII solutions. 

\item The S-V and S-VII solutions can lead to more than 25\% enhancement in 
 $R^{\tau \mu}_{A_{FB}}$ over the SM value. No class-B solutions can induce large new physics effects in this observable. 

\item Amongst the flavor ratio of optimized observables, $R^{\tau \mu}_{P_2}$ and $R^{\tau \mu}_{P'_5}$ can show maximum deviation, up to 25\%,  from the SM. This deviation is possible for new physics scenario S-V. For other ratios, only marginal deviation is allowed. 
\end{itemize}
Therefore, in the considered framework, the current $b \to s \mu^+ \mu^-$  data does allow for large new physics effects in a number of $B \to K^* \tau^+ \tau^-$ observables. These effects range from  20\% - 30\% up to an order of magnitude above the SM level. Hence $B \to K^* \tau^+ \tau^-$ decay mode has immense potential to probe physics beyond SM, particularly by  complementing the quest for new physics signatures in  $b \to s \mu^+ \mu^-$  sector.

\bigskip
\noindent
{\bf Acknowledgements}: I would like to thank David London for his helpful advice and useful comments on the project. I also  thank  Shireen Gangal for  fruitful discussions. I am also thankful to  Ashutosh Kumar Alok for his useful suggestions and discussions along with corrections on the manuscript and Arindam Mandal for critical reading of the manuscript. 

\appendix
\section{Angular coefficients}
\label{appen}
The twelve $q^2$ dependent angular coefficients $I^{(a)}_i$  in eq~(\ref{Ifunc}) can be expressed in terms of transversity amplitudes which are given by~\cite{Altmannshofer:2008dz}
\begin{eqnarray}
I_1^s &=& \frac{(2+\beta^2_{\ell})}{4}\left[|A^L_{\perp}|^2+|A^L_{\parallel}|^2 +(L\to R)\right] \nonumber \\
&&+ \frac{4m^2_{\ell}}{q^2} {\rm Re}\left(A^L_{\perp}A^{R*}_{\perp}+A^L_{\parallel}A^{R*}_{\parallel}\right), \nonumber 
\end{eqnarray}
\begin{eqnarray}
I^c_1 & = & |A^L_{0}|^2+|A^R_{0}|^2 \nonumber \\
&&+\frac{4m^2_{\ell}}{q^2}\left[|A_t|^2 + 2 {\rm Re}\left(A^L_0 A^{R*}_0\right)\right] +\beta^2_{\ell} |A_S|^2,\nonumber 
\end{eqnarray}
\begin{eqnarray}
I_2^s &=& \frac{\beta^2_{\ell}}{4}\left[|A^L_{\perp}|^2+|A^L_{\parallel}|^2 + (L\to R)\right],\nonumber
\end{eqnarray}
\begin{eqnarray}
I^c_2 &= & -\beta^2_{\ell} \left[|A^L_0|^2 + |A^R_0|^2\right], \nonumber 
\end{eqnarray}
\begin{eqnarray}
I_3 &= & \frac{\beta^2_{\ell}}{2} \left[|A^L_{\perp}|^2 - |A^L_{\parallel}|^2 + (L\to R)\right], \nonumber
\end{eqnarray}
\begin{eqnarray}
I_4 &  = & \frac{\beta^2_{\ell}}{\sqrt{2}} \left[ {\rm Re}(A^L_0 A^{L*}_{\parallel}) + (L\to R)\right],\nonumber
\end{eqnarray}
\begin{eqnarray}
I_5/\sqrt{2} \beta_{\ell} &= &{\rm Re}(A^L_0 A^{L*}_{\perp}) -(L\to R)
\nonumber \\
&&- \frac{m_{\ell}}{\sqrt{q^2}} {\rm Re}(A^L_{\parallel}A^*_S+A^R_{\parallel}A^*_S),\nonumber 
\end{eqnarray}
\begin{eqnarray}
I^s_6 &= & 2\beta_{\ell}\left[{\rm Re}(A^L_{\parallel}A^{L*}_{\perp})- (L\to R)\right],\nonumber
\end{eqnarray}
\begin{eqnarray}
I^c_6 &=& 4\beta_{\ell}\frac{m_{\mu}}{\sqrt{q^2}}{\rm Re}\left[A^L_0A^*_S + (L\to R)\right],\nonumber
\end{eqnarray}
\begin{eqnarray}
I_7/\sqrt{2}\beta_{\ell} & =& {\rm Im}(A^L_0 A^{L*}_{\parallel})- (L\to R)\nonumber \\
&& +\frac{m_{\mu}}{\sqrt{q^2}}{\rm Im}(A^L_{\perp}A^*_S + A^R_{\perp}A^*_S),\nonumber
\end{eqnarray}
\begin{eqnarray}
I_8 &= & \frac{\beta^2_{\ell}}{\sqrt{2}}\left[{\rm Im}(A^L_0A^{L*}_{\perp}) + (L\to R)\right],\nonumber
\end{eqnarray}
\begin{eqnarray}
I_9 &= & \beta^2_{\ell}\left[{\rm Im}(A^{L*}_{\parallel}A^L_{\perp})+ (L\to R)\right].
\end{eqnarray}
The expression of transversity amplitudes which are written in terms of form factors  $V(q^2)$, $A_{0,1,2}(q^2)$ and $T_{1,2,3}(q^2)$ can be found in ref.~\cite{Bharucha:2015bzk}.

\end{document}